\documentclass[a4paper,11pt]{article}
\pdfoutput=1 

\usepackage{jheppub,bm} 

\usepackage[T1]{fontenc} 

\newcommand{\nn}{\nonumber}

\def\bI{\boldsymbol{I}}

\def\beq{\begin{equation}}
\def\eeq{\end{equation}}
\def\beqn{\begin{eqnarray}}
\def\eeqn{\end{eqnarray}}

\def\spa#1.#2{\left\langle#1#2\right\rangle}
\def\spab#1.#2.#3{\left\langle#1|#2|#3\right]}
\def\spb#1.#2{\left[#1#2\right]}

\allowdisplaybreaks[4]

\begin{document}

\title{Predictions for diphoton production at the LHC through NNLO in QCD}

\author[a]{John M. Campbell}
\author[b]{, R. Keith Ellis}
\author[a]{, Ye Li}
\author[c]{and Ciaran Williams}

\affiliation[a]{Fermilab,\\PO Box 500, Batavia, IL 60510, USA}
\affiliation[b]{Institute for Particle Physics Phenomenology,\\Department of Physics, Durham University, Durham DH1 3LE, United Kingdom}
\affiliation[c]{Department of Physics,\\ University at Buffalo, The State University of New York, Buffalo
14260, USA}

\emailAdd{johnmc@fnal.gov}
\emailAdd{keith.ellis@durham.ac.uk}
\emailAdd{yli32@fnal.gov}
\emailAdd{ciaranwi@buffalo.edu}

\newcommand{\zero}{{(0)}}
\newcommand{\one}{{(1)}}
\newcommand{\two}{{(2)}}
\newcommand{\ztwo}{\zeta_2}
\newcommand{\zthree}{\zeta_3}
\newcommand{\cf}{C_F}
\newcommand{\ca}{C_A}
\newcommand{\nf}{n_f}
\newcommand{\cfs}{C_F^2}
\newcommand{\Tcm}{\tau_{cm}}

\newcommand{\DeltaNNNLO}{\Delta \sigma^{{\rm{N3LO}}}_{gg,n_F}}

\abstract{
In this paper we present a next-to-next-to-leading order  (NNLO) calculation of the process $pp\rightarrow \gamma\gamma$
that we have implemented into the parton level Monte Carlo code MCFM.
We do not find agreement with the previous calculation of this process in the literature.
In addition to the  $\mathcal{O}(\alpha_s^2)$ corrections present at NNLO, we include some effects arising at $\mathcal{O}(\alpha_s^3)$,
namely those associated with gluon-initiated closed fermion loops. 
We investigate the role of this process in the context of studies of QCD at colliders and as a background for searches for new physics,
paying particular attention to the diphoton invariant mass spectrum.  We demonstrate that the NNLO QCD prediction for the shape of this
spectrum agrees well with functional forms used in recent data-driven fits.}

\preprint{
\noindent IPPP/16/16 \\
\hspace*{\fill} FERMILAB-PUB-16-074-T}
\maketitle
\flushbottom
\section{Introduction}

The discovery of a light Higgs boson~\cite{Chatrchyan:2012ufa,Aad:2012tfa}, which decays to two photons, 
has helped cement the diphoton process as one of the most interesting final states to study during the second run of 
the LHC (Run II).  Experimental studies of prompt ($\gamma$) and diphoton $(\gamma\gamma)$ production at hadron colliders 
have been undertaken for several decades~\cite{Abachi:1996qz,Bonvin:1988yu,Albajar:1988im,Alitti:1992hn,Abe:1992cy,Abazov:2010ah,Aaltonen:2012jd,Chatrchyan:2011qt,Aad:2012tba,Aad:2013zba,Aaltonen:2011vk,Chatrchyan:2013mwa,Chatrchyan:2014fsa}. 
These studies are possible in part due to the high rate of production, but also because of the relative cleanliness of the experimental final state.
As the energy available for collisions has increased, and as the amount of data collected has grown, so too has the
region of diphoton invariant mass ($m_{\gamma\gamma}$) that can be probed.
At the LHC experimental data is now available up to scales of order $1$~TeV,  allowing 
for searches for new heavy resonances that may decay to photon pairs~\cite{ATLAS-CONF-2015-081,CMS-PAS-EXO-15-004}. 

During Run II, the large data set will result in many measurements being performed at a level of detail that demands exquisite theoretical predictions.
Therefore, in addition to the detailed experimental studies, the prompt and diphoton processes have received considerable theoretical attention.  The next-to-leading order (NLO) calculations embodied
in the Jetphox~\cite{Catani:2002ny} and Diphox~\cite{Binoth:1999qq} Monte Carlo codes have  been extensively utilized in
the experimental literature. In addition to the NLO calculation of diphoton production, $gg\rightarrow \gamma\gamma$ contributions that are formally
higher-order, but phenomenologically important, have also been computed~\cite{Bern:2001df,Bern:2002jx}.
However, existing 7~TeV analyses have already confirmed the inadequacies of NLO calculations when confronted with
data~\cite{Aad:2012tba,Chatrchyan:2014fsa}.  Instead, much better agreement is found with the
recently-completed next-to-next-to Leading Order (NNLO) calculation~\cite{Catani:2011qz}, that naturally subsumes the first $gg\rightarrow \gamma\gamma$ contributions.

This calculation was made possible through the application of the $Q_T$-subtraction procedure~\cite{Catani:2007vq}. This procedure makes use of the known factorization
properties at small transverse momenta of the diphoton system to efficiently handle complications arising from infrared singularities.   Although a variety
of other methods for regularizing and combining infrared singularities have been devised~\cite{GehrmannDeRidder:2005cm,Czakon:2010td,Cacciari:2015jma}, and used to
provide a suite of new predictions for $2\rightarrow 2$ hadron collider
processes~\cite{Boughezal:2013uia,Currie:2013dwa,Brucherseifer:2014ama,Chen:2014gva,Ridder:2015dxa,Czakon:2015owf},
the relative simplicity of the $Q_T$-subtraction method is highly appealing~\cite{Catani:2009sm,Ferrera:2011bk,Grazzini:2013bna,Cascioli:2014yka,Gehrmann:2014fva,Grazzini:2015nwa,Grazzini:2015hta}.
The $Q_T$-subtraction method generates a counter-term that regularizes the singularity as $Q_T \to 0$ but is otherwise non-local;  it also naturally lends itself to implementation
as a slicing method (``$Q_T$-slicing'').
A promising new development is a generalization of the $Q_T$-based methods, which were originally only applicable to color-neutral final states,
to new methods~\cite{Gao:2012ja,Boughezal:2015dva,Gaunt:2015pea} based on Soft Collinear Effective Field Theory
(SCET)~\cite{Bauer:2000ew,Bauer:2000yr,Bauer:2001yt,Bauer:2001ct,Bauer:2002nz}.    One of these methods~\cite{Boughezal:2015dva,Gaunt:2015pea},
based on the $N$-jettiness global event shape~\cite{Stewart:2010tn}, can in principle be applied to arbitrary
processes~\cite{Boughezal:2015dva,Boughezal:2015dra,Boughezal:2015aha,Boughezal:2015ded,Campbell:2016jau}.
In its implementation as a slicing method, the $N$-jettiness variable ($\tau$) is used to split the phase space into two regions. In the region where $\tau > \tau^{\rm{cut}}$ at
least one of the additional partons is resolved. Therefore the calculation contains only single unresolved limits and is amenable to calculation using standard NLO techniques. For the
second region, where  $\tau  < \tau^{\rm{cut}}$, both partons can be simultaneously unresolved. In this region a factorization theorem 
from SCET~\cite{Stewart:2010tn} is used to approximate the
cross section to the desired perturbative accuracy.  This is a natural generalization of $Q_T$-subtraction, where a similar reasoning applies when replacing $\tau$ with $Q_T$
and SCET factorization with one based on the Collins-Soper-Sterman formalism~\cite{Collins:1984kg}.

The aim of this paper is to present a new NNLO calculation of $pp\rightarrow \gamma\gamma$ using the $N$-jettiness slicing approach and compare it
with the existing calculation of ref.~\cite{Catani:2011qz}.  Given its importance for Run II phenomenology an
independent calculation is crucial. In fact we will find that we cannot reproduce the results of the literature and we believe that existing results for
this process are inaccurate.  This underlines the need for multiple independent calculations of processes such as this one that are of great importance
for existing and future experimental analyses.  We investigate the role of higher-order effects to the $gg$ initiated closed loops of
quarks, and combine this prediction with NNLO for the first time. We will also investigate the role of top quark loops at high invariant masses.  Our calculation is
implemented in MCFM~\cite{Campbell:1999ah,Campbell:2011bn,Campbell:2015qma} and will be released in a forthcoming version of the code.

We continue this paper by outlining the various component pieces of our calculation in section~\ref{sec:calc}. In section~\ref{sec:valid} we compare
our predictions to existing results from the literature and discuss the checks we performed on our calculation.
In section~\ref{sec:pheno} we turn our attention to LHC phenomenology, comparing our predictions to data obtained by the CMS experiment at $7$~TeV,
and to the $m_{\gamma\gamma}$ spectrum reported by ATLAS at $13$~TeV. Finally, we draw our conclusions in section~\ref{sec:conc}. Appendices~\ref{app:cal},~\ref{app:top}  and~\ref{app:nfrat} contain additional technical details of our calculation.

\section{Calculation} 
\label{sec:calc}

In this section we present an overview of our calculation of diphoton production at NNLO and discuss the various contributions 
that are included in this paper.  Before going into detail we introduce the following notation  
\begin{eqnarray}
\label{eq:orderdef}
\sigma^{NLO}_{\gamma\gamma} &=&   \sigma^{LO} + \Delta \sigma^{NLO} \nn \, ,\\
\sigma^{NNLO}_{\gamma\gamma}  &=& \sigma^{NLO} + \Delta \sigma^{NNLO} =  \sigma^{LO} + \Delta \sigma^{NLO}  + \Delta \sigma^{NNLO} \, .
\end{eqnarray}
In this way $\Delta \sigma^{X}$ represents the correction obtained from including the coefficient that first arises at order
$X$ in perturbation theory. 
We use this notation both inclusively (as written above) and for differential predictions. 

\subsection{Overview}

\begin{figure}
\begin{center}
\includegraphics[width=15cm]{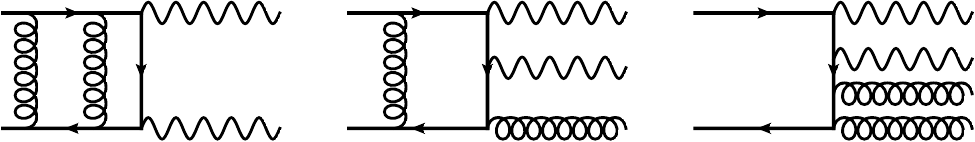}
\caption{Representative Feynman diagrams for the calculation of $pp\rightarrow \gamma\gamma$ at NNLO.
From left to right these correspond to double virtual (calculated in ref.~\cite{Anastasiou:2002zn}),
real-virtual and real-real corrections.}
\label{fig:gagaFD}
\end{center}
\end{figure}
We present representative Feynman diagrams for the various topologies that enter the calculation of the $pp\rightarrow \gamma\gamma$ process at NNLO in
Figure~\ref{fig:gagaFD}. At this order in perturbation theory contributions arise from three distinct final states. The simplest is
the one that also represents the Born contribution and corresponds to a $2\rightarrow 2$ phase space. At NNLO this final state
receives corrections from two-loop amplitudes interfered with the LO amplitude, and one-loop squared contributions. The $2\rightarrow
3$ real-virtual phase space consists of tree-level and one-loop amplitudes for $q\overline{q}g \gamma\gamma$ interfered with one
another. Finally the largest phase space, representing a $2\rightarrow 4$ process, is referred to as the double-real contribution
and consists of two tree-level $q\overline{q} \gamma\gamma+2$ parton amplitudes squared.   The contributions discussed above have
ultraviolet (UV) poles in the double-virtual and real-virtual phase spaces, which we renormalize in the $\overline{\rm{MS}}$ scheme.
Amplitudes for the double-virtual contribution can be found in ref.~\cite{Anastasiou:2002zn}, for the real-virtual
in ref.~\cite{Campbell:2014yka}, and tree-level amplitudes for the real-real can be found in ref.~\cite{DelDuca:1999pa}.

After UV renormalization the individual component pieces of the calculation still contain singularities of infrared (IR)
origin.  These infrared poles must be regulated, made manifest, and combined across the different phase spaces in order to ensure that a sensible
prediction is obtained.  As discussed in the introduction, we will use the $N$-jettiness slicing
technique proposed in refs~\cite{Gaunt:2015pea,Boughezal:2015dva} for this task.  This results in an above-cut contribution
corresponding to the calculation of $pp\rightarrow \gamma\gamma j$ at NLO.  The below-cut contribution requires 
2-loop soft~\cite{Kelley:2011ng,Monni:2011gb} and beam~\cite{Gaunt:2014xga} functions,
together with the process-dependent hard function.  Various component pieces of this
calculation, including explicit results for the hard function, are given in Appendix~\ref{app:cal}

\subsection{$gg$ initiated loops at LO and NLO}
\label{subsec:gg}

\begin{figure}
\begin{center}
\includegraphics[width=10cm]{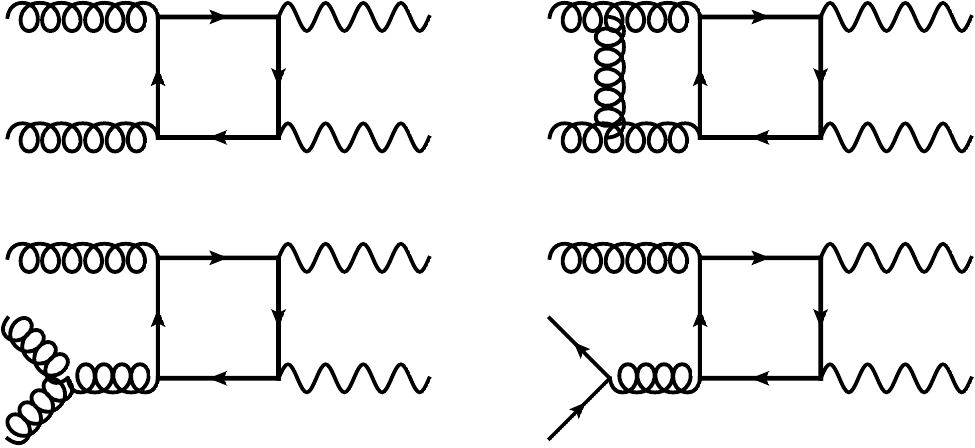}
\caption{Representative Feynman diagrams for the calculation of $gg\rightarrow \gamma\gamma$ at LO (top left) and NLO (the remainder). 
The virtual two-loop corrections are shown in the top right, while the bottom row corresponds to real radiation contributions. }
\label{fig:gagaFDgg}
\end{center}
\end{figure}
The NNLO calculation of $\gamma\gamma$ production represents the first order in perturbation theory that is 
sensitive to $gg$ initial states.  One class of $gg$ configurations corresponds to real-real corrections,
i.e. the $gg\rightarrow q\overline{q} \gamma\gamma$ matrix element that is related to the contribution shown in
figure~\ref{fig:gagaFD} (right) by crossing. These pieces are combined with contributions from the DGLAP evolution of the parton 
distribution functions in the real-virtual and double-virtual terms to ensure an IR-finite result.
The second type of contribution is due to $n_F$ ``box'' loops, for which a representative  Feynman diagram 
is shown  in the top left corner of Figure~\ref{fig:gagaFDgg}.  This contribution has no tree-level analogue and
is thus separately finite.

The box diagrams result in a sizeable cross section ($\approx \sigma_{LO}$), primarily due to the large gluon flux at LHC energies
and the fact that this contribution sums over different quark flavors in the loop. In this section, we focus on $n_F=5$ light quark
loops.  Since this contribution is clearly important for phenomenology it is interesting to try to isolate and compute higher order
corrections to it.  We illustrate typical component pieces of these NLO corrections in the remaining diagrams in
Figure~\ref{fig:gagaFDgg}. They comprise two-loop $gg \gamma\gamma$  amplitudes, and one-loop $ggg\gamma\gamma$ and
$gq\overline{q}\gamma\gamma$ amplitudes.  A NLO calculation of  $gg\rightarrow \gamma\gamma$ including the two-loop and one-loop
$ggg\gamma\gamma$ amplitudes was presented in refs.~\cite{Bern:2001df,Bern:2002jx}.  An infrared-finite calculation can be obtained
from the $gg\rightarrow\gamma\gamma$ two-loop amplitudes and the $ggg\gamma\gamma$ one-loop amplitudes, provided that a suitable
modification to the quark PDFs is used (essentially using a LO evolution for the quark PDFs and a NLO evolution for the gluon PDFs).
On the other hand if the  $q\overline{q}g\gamma\gamma$ amplitudes are included then the corresponding collinear singularity can be
absorbed into the quark PDFs as normal at NLO, allowing for a fully consistent treatment. In the original
calculation~\cite{Bern:2001df,Bern:2002jx} (and the corresponding implementation in MCFM~\cite{Campbell:2011bn}) the first approach
was taken. Here we will follow the second approach and include the $q\overline{q}g\gamma\gamma$ amplitudes.  Although formally
an improvement, we find that the differences between the two approaches are negligible. Most of the required $q\overline{q}g\gamma\gamma$ amplitudes can be found
in ref.~\cite{Campbell:2014yka}.  However, since that paper was concerned only with the NLO predictions for the $\gamma\gamma j$ process, it did not include the one-loop amplitude
that interferes with a vanishing tree-level term.  In the calculation presented here this purely-rational amplitude is squared and therefore must be
properly included. For completeness we present this missing amplitude in Appendix~\ref{app:nfrat}.

Since the NLO corrections to the $gg$ initiated diagrams form a part of the N$^{3}$LO cross section but do not represent a full calculation at that order,
we define the additional cross section associated with them as $\DeltaNNNLO$. The subscript indicates that they are associated
with $gg$ initiated closed loops of quarks.  Although by no means a complete $\mathcal{O}(\alpha_s^3)$ prediction,  it is possible that the
$\DeltaNNNLO$ contribution forms a sizeable part of this correction. The impact of these terms will be discussed at length in
section~\ref{sec:pheno}.

\subsection{Impact of the top quark at high $m_{\gamma\gamma}$} 
\label{sec:toploops}

\begin{figure}
\begin{center}
\includegraphics[width=0.75\textwidth]{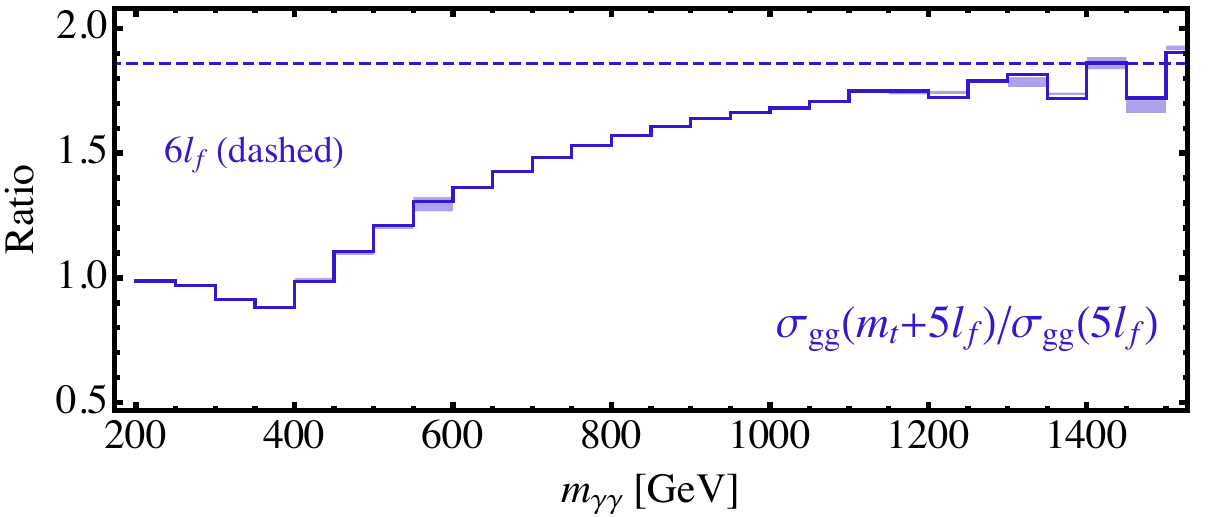}
\caption{The ratio of the invariant mass distribution in $gg \to \gamma\gamma$ computed using five light flavors and the effect of the
top quark, to the calculation with $n_F=5$ alone.  The dashed line shows the ratio of the result for $n_F=6$ to the one for $n_F=5$ that
corresponds to Eq.~(\ref{eq:nf6tonf5}).}
\label{fig:ggmt_mgg}
\end{center}
\end{figure} 
The previous subsection outlined the calculation of $gg$ loops for $n_F=5$ light quarks.
While this is an excellent approximation for low invariant mass photon
pairs,  at higher energies this is no longer appropriate due to contributions from top quarks circulating in the loop.
Current searches for physics beyond the Standard Model are sensitive to regions of large invariant mass
$m_{\gamma\gamma} > 2m_t$, so it is essential to quantify the role of the top quark in this region of phase space. This is the primary aim of this
section.  To that end we have computed the amplitudes for $gg\rightarrow \gamma\gamma$ that proceed through a closed loop of heavy quarks, and 
include details of the calculation in Appendix~\ref{app:top}. 

We will perform a detailed phenomenological study of the high invariant mass region in section~\ref{sec:pheno}, but to illustrate the importance of the
top quark loop we assess its impact on the relevant invariant mass spectrum in Figure~\ref{fig:ggmt_mgg}.  The results have been obtained for
the LHC operating at $13$~TeV and under fiducial cuts inspired by the ATLAS collaboration~\cite{ATLAS-CONF-2015-081} that are described in
section~\ref{sec:pheno}.  We show the ratio of the result with $n_F=5$ light flavors (for the $gg$ initiated pieces only) and the top quark loop included, to the result for $n_F=5$ light
flavors alone. There is a slight decrease in the prediction below the $2m_t$ threshold, due to the effects of a destructive interference, then a steady
rise to an asymptotic value.  This asymptotic value is of course the result for $n_F=6$ light quark flavors (without including any modification to the
running of $\alpha_s$) and is simply given by,
\begin{eqnarray}
\label{eq:nf6tonf5}
\frac{\sigma_{gg}(n_F=6)}{\sigma_{gg}(n_F=5)} = \left(\frac{3 Q_u^2 + 3 Q_d^2}{2 Q_u^2+3Q_d^2}\right)^2 = 1.8595\dots
\end{eqnarray}

\subsection{Summary} 

In this section we have presented an overview of the various component pieces of our calculation. For the bulk of this paper we will define our NNLO
calculation to only account for five light flavors of quarks. Unless otherwise stated we do not include the NLO corrections to the $gg$ initial state
that have been discussed in section~\ref{subsec:gg}. Instead we refer
to these pieces always as  $\sigma^{\rm{NNLO}}+\DeltaNNNLO$. 
Our default scale choice for the renormalization and factorization scales will
be $\mu=m_{\gamma\gamma}$. We estimate the theoretical uncertainty by varying
this central scale by a factor of two in each direction, i.e.
$\mu=2m_{\gamma\gamma}$ and $\mu=0.5 m_{\gamma\gamma}$. This variation will be
indicated by shaded bands in the figures of Section~\ref{sec:pheno}.

\section{Validation} 
\label{sec:valid}

In this section we compare our results for $pp \rightarrow \gamma\gamma$ with those presented in ref.~\cite{Catani:2011qz}.
A summary of cross-sections that have been computed in that work is shown in Table~\ref{tab:CG1}.
To emulate their calculation we impose a series of phase space selection cuts. 
The cuts on the transverse momenta of the photons depend on their relative size, 
$p_T^{\rm{hard}} > 40$~GeV and $p_T^{\rm{soft}} > 25$~GeV.
The photons are also required to be central, $|\eta_{\gamma} | < 2.5$ 
and in addition we require that the invariant mass of the photon-photon 
system lies in the interval $20  \le m_{\gamma\gamma} \le 250$~GeV.
Finally at NLO and NNLO we impose the following isolation requirement~\cite{Frixione:1998jh} 
\begin{eqnarray} 
E_T^{had}(r) \le \epsilon_{\gamma} p^{T}_\gamma\left(\frac{1-\cos{r}}{1-\cos{R}}\right)^n  \, ,
\end{eqnarray} 
with $n=1$, $\epsilon_{\gamma} =0.5$ and  $R= 0.4$. 
We use $\alpha = 1/137$ and the remaining EW parameters are set to the default values in MCFM.
The PDFs are taken from MSTW2008~\cite{Martin:2009iq} and are matched to the appropriate order in perturbation theory. The
renormalization and factorization scales are mostly set to the invariant mass of the photon pair $\mu_F=\mu_F=m_{\gamma\gamma}$, although
we will also present results for $\mu_F=\mu_R=m_{\gamma\gamma}/2$ and $\mu_F=\mu_R=2 m_{\gamma\gamma}$. 

The results that we obtain from our implementation in MCFM are presented
in Table~\ref{tab:MCFM} and should be compared with the results from ref.~\cite{Catani:2011qz}
that are shown in Table~\ref{tab:CG1}.
Whilst our LO and NLO predictions are in good accord, we find no such agreement for the 
NNLO cross sections, for any of the choices of scale. The discrepancy is approximately $3$pb, or around $8$\% of the total NNLO prediction.
However we do note that the size of the scale variation,
i.e. the departures from the central choice, is the same for both calculations.
\begin{table} 
\begin{center}
\begin{tabular}{|c|c|c|c|} 
\hline
$\sigma $[fb] & LO & NLO & NNLO \\
\hline
\hline
$\mu_F =\mu_R = m_{\gamma\gamma}/2  $ &  5045 $\pm$ 1 & 26581 $\pm$ 23 & 45588 $\pm$ 97   \\
\hline
$\mu_F =\mu_R = m_{\gamma\gamma}$ &  5712 $\pm$ 2 & 26402 $\pm$ 25 &  43315 $\pm$ 54 \\
\hline
$\mu_F =\mu_R = 2m_{\gamma\gamma} $ & 6319 $\pm$ 2 & 26045 $\pm$ 24 & 41794 $\pm$ 77  \\
\hline\hline
\end{tabular} 
\caption{Cross sections reported in ref.~\cite{Catani:2011qz}.}
\label{tab:CG1}
\end{center}
\end{table}
\begin{table} 
\begin{center}
\begin{tabular}{|c|c|c|c|} 
\hline
$\sigma $[fb] & LO & NLO & NNLO \\
\hline
\hline
$\mu_F =\mu_R = m_{\gamma\gamma}/2  $ &  5043 $\pm$ 1 & 26578 $\pm$ 13 & 42685 $\pm$ 35  \\
\hline
$\mu_F =\mu_R = m_{\gamma\gamma}$ &  5710 $\pm$ 1 & 26444 $\pm$ 12 &  40453 $\pm$ 30 \\
\hline
$\mu_F =\mu_R = 2m_{\gamma\gamma} $ & 6315 $\pm$ 2 & 26110 $\pm$ 13 &  38842 $\pm$ 27  \\
\hline\hline
\end{tabular} 
\caption{Cross section results obtained using MCFM.  The NLO contribution is always computed using
Catani-Seymour dipole subtraction;  the  NNLO coefficient corresponds to the $\tau \rightarrow 0$ limit
of a calculation using $N$-jettiness regularization (c.f. Figure~\ref{fig:nnlotaudep}).  In the NNLO
calculation the errors are obtained by adding the fitting and NLO Monte Carlo uncertainties in quadrature.}
\label{tab:MCFM}
\end{center}
\end{table}

Since we therefore do not agree with the essential results of the existing literature we now describe the further checks that we
have performed on our calculation. Several of the ingredients for the below-cut contribution have been reused from previous
calculations where good agreement with the literature results was obtained. Specifically, the soft and beam functions have already
been used to compute the Drell-Yan and associated Higgs production processes~\cite{Boughezal:2016wmq,Campbell:2016jau}.  The MCFM
predictions for these cross sections agree perfectly with the known results from the literature.  The remaining below-cut
contribution, the hard function, has been implemented in two independent codes that check both the SCET matching and the
proper inclusion of the double-virtual results of ref.~\cite{Anastasiou:2002zn}\footnote{We have adjusted the results of ref.~\cite{Anastasiou:2002zn} 
to account for small typos in the manuscript, as detailed in Appendix~\ref{app:cal}.}. 
 Additionally we have checked that by setting $\mu^2 =s$, and implementing the hard
function for a specific scale, we can reproduce the full result by application of the renormalization group equations. This test is
extremely non-trivial since the $\mu^2$ dependence occurs both in the finite functions taken from ref.~\cite{Anastasiou:2002zn} (in
their notation, a dependence on $S$) and also in the matching to the SCET formalism. This check therefore ensures that no mistakes are
made in the relative normalization between the two parts of the hard function calculation. For the $gg\rightarrow \gamma\gamma$
pieces we have reproduced the results of refs.~\cite{Bern:2001df,Bern:2002jx}, which were implemented previously in
MCFM~\cite{Campbell:2011bn}. For the above-cut pieces we have compared our NLO prediction for $\gamma\gamma j$ with the results
presented in ref.~\cite{Gehrmann:2013aga}, finding agreement for the isolation procedure used here (``smooth-cone''). We have also checked the analytic
calculation of the helicity amplitudes for the real and virtual contributions to $\gamma\gamma j$ production against
an in-house implementation of the numerical $D$-dimensional algorithm~\cite{Ellis:2008ir}.

In order to eliminate the $N$-jettiness slicing procedure as a cause of the difference, we have also implemented
$Q_T$-slicing in MCFM.\footnote{The $Q_T$-slicing method is based on the same factorization and ingredients that were used in the
previous $Q_T$-subtraction calculation~\cite{Catani:2007vq}.}
  This implementation has been additionally checked, for large values of $Q_T^{\rm{cut}}$, with a
calculation using a completely different setup.  The alternate $Q_T$-slicing calculation is implemented using
the Sherpa framework~\cite{Gleisberg:2008ta} and uses the OpenLoops~\cite{Cascioli:2011va} and
BlackHat~\cite{Berger:2008sj,Bern:2011pa,Bern:2014vza} programs to evaluate the above-cut
matrix elements.   An obvious cause for concern in either of these slicing-based methods is the dependence on the regulating
parameter.  When comparing our predictions it is therefore crucial to investigate the dependence of them on this unphysical
slicing parameter, either $\tau^{\rm{cut}}$ or $Q_T^{\rm{cut}}$ as appropriate.

\begin{figure}
\begin{center}
\includegraphics[width=0.8\textwidth]{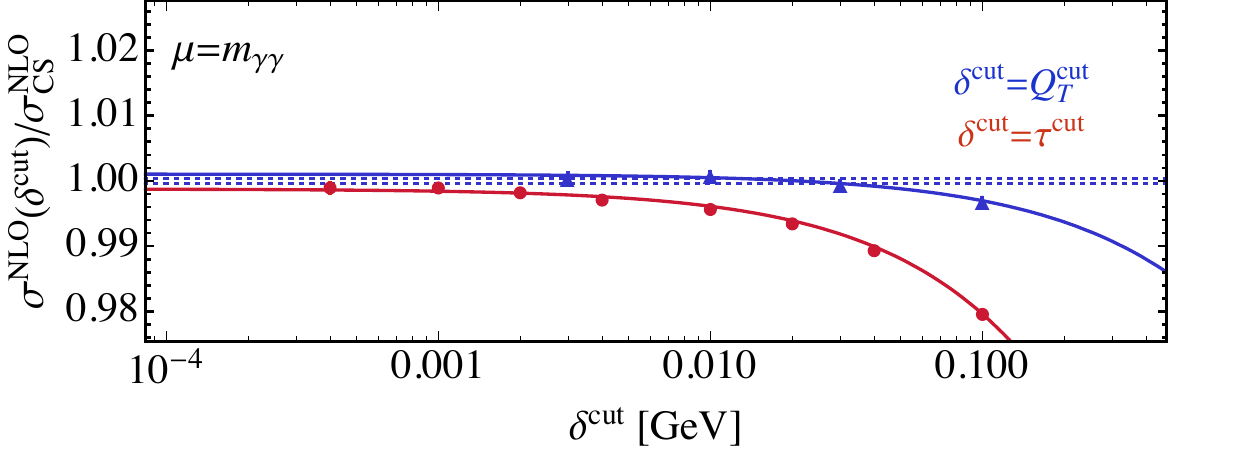}
\caption{The dependence of the NLO cross section on the slicing parameter $\delta^{\rm{cut}}$.  Results are presented
using the $N$-jettiness (circles) ($\delta^{\rm{cut}} \equiv \tau^{\rm{cut}}$) and $Q_T$-slicing (triangles)
($\delta^{\rm{cut}} \equiv Q_T^{\rm{cut}}$) methods.  In both cases the results are normalized to
the standard MCFM prediction obtained with Catani-Seymour dipole subtraction, which does not have a slicing parameter dependence. }
\label{fig:nlotaudep}
\end{center}
\end{figure} 
As a point of reference, we first study the dependence of the total NLO cross section on the slicing parameter
in Figure~\ref{fig:nlotaudep}.  To assess the agreement with the known result, we divide the results of
these calculations with the one obtained from the existing NLO calculation of MCFM.  This implementation of the
$pp \to \gamma\gamma$ process~\cite{Campbell:2011bn} uses Catani-Seymour dipoles~\cite{Catani:1996vz} to regulate the
infrared divergences and thus contains no dependence on a slicing parameter. The figure indicates that the 
slicing results approach the correct cross section, with deviations in the cross section that are $\mathcal{O}(0.1)$\% and smaller
for $\tau^{\rm{cut}} \lesssim 0.002$~GeV or $Q_T^{\rm{cut}} \lesssim 0.04$~GeV. This agreement is an additional
check of the correctness of the NNLO calculation since the one-loop hard function is also used there.

Although the effect of power corrections appears to
be milder for $Q_T$-slicing than $N$-jettiness regularization, by around a factor of 20, we note that the computational
resources required to perform the calculations at these two points is similar.   The resources needed for a computation of a given
accuracy is dominated by the calculation of the above-cut contribution, which scales as~\cite{Berger:2010xi,Gaunt:2015pea},
\begin{equation}
\Delta\sigma^{N^nLO}(\tau > \tau^{\rm{cut}}) / \sigma^{LO} \sim \frac{1}{n!} \left(\frac{\alpha_s C_F}{\pi}\right)^n \log^{2n} \frac{\tau^{\rm{cut}}}{Q} + \ldots
\end{equation}
for the $N$-jettiness calculation.  In this equation $Q$ is an appropriate hard scale that is given here by the transverse momentum
of the photons.  A similar analysis for $Q_T$-slicing yields the result~\cite{Catani:2001cr},
\begin{equation}
\Delta\sigma^{N^nLO}(Q_T > Q_T^{\rm{cut}}) / \sigma^{LO} \sim  \frac{1}{n!} \left(\frac{2 \alpha_s C_F}{\pi}\right)^n \log^{2n} \frac{Q_T^{\rm{cut}}}{Q} + \ldots
\end{equation}
Therefore one expects similar computational effort for the two methods when the values of $\tau^{\rm{cut}}$ and $Q_T^{\rm{cut}}$
are related by~\cite{Berger:2010xi},
\begin{equation}
\label{eq:tauequivQT}
\frac{\tau^{\rm{cut}}}{Q} \simeq \left(  \frac{Q_T^{\rm{cut}}}{Q} \right)^{\sqrt{2}} \,.
\end{equation}
For $Q=40$~GeV one therefore expects the NLO calculation using $Q_T^{\rm{cut}}=0.04$~GeV to be as
expensive as the one with $\tau^{\rm{cut}} = 0.0023$~GeV.

\begin{figure}
\begin{center}
\includegraphics[width=0.8\textwidth]{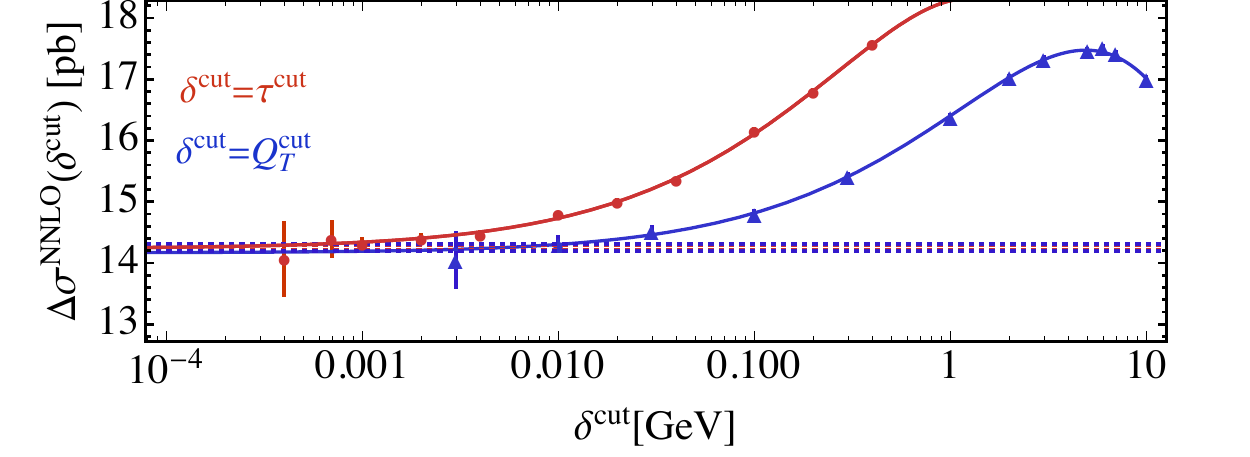}
\caption{The dependence of the NNLO coefficient $\Delta  \sigma^{\rm{NNLO}}$ on the slicing parameter $\delta^{\rm{cut}}$.
Results are presented using the $N$-jettiness ($\delta^{\rm{cut}} \equiv \tau^{\rm{cut}}$) (circle) and $Q_T$-slicing (triangles)
($\delta^{\rm{cut}} \equiv Q_T^{\rm{cut}}$) methods. 
The dashed lines correspond to the errors associated with the fitting procedure.}
\label{fig:nnlotaudep}
\end{center}
\end{figure} 
Figure~\ref{fig:nnlotaudep} shows the $\delta^{\rm{cut}}$ dependence  for the NNLO coefficient, $\Delta \sigma^{NNLO}$
(c.f. Eq.~\ref{eq:orderdef}). It is clear that the dependence is much more pronounced than at NLO.
To achieve a 1\% accuracy for $\Delta \sigma^{NNLO}$ requires a value of $\tau^{\rm{cut}}$
around $0.002$~GeV or $Q_T^{\rm{cut}}$ smaller than about $0.02$~GeV. 
Once again power corrections are less significant for $Q_T$-slicing, but the computing time to achieve equivalent  accuracy is
comparable in both methods.  This is in line with the scaling expected from Eq.~(\ref{eq:tauequivQT}).
The NNLO results reported in Table~\ref{tab:MCFM} are obtained from the asymptotic $\tau\rightarrow 0$
results obtained by a fit to the $\tau^{\rm{cut}}$ dependence that is represented by the solid red line in
figure~\ref{fig:nnlotaudep}. We observe that for values of $Q_T^{\rm{cut}}$ around $1$~GeV there is a a local maximum in the NNLO coefficient, 
which could be mistaken for the onset of asymptotic
behavior.

We have communicated our findings with the authors of ref.~\cite{Catani:2011qz}, who have acknowledged a problem with their results presented in ref~\cite{Catani:2011qz}. The updated version of their code produces results that are consistent with ours, within Monte Carlo
 uncertainties.

\section{LHC Phenomenology} 
\label{sec:pheno}

In this section we present results that are relevant for current LHC phenomenology.
We first investigate the comparison of our calculation with existing data taken by the CMS experiment with the LHC
operating at $\sqrt{s}=7$~TeV.  Although such comparisons have already been performed, we believe that this is especially important given the disagreement
with the previous NNLO calculation noted in section~\ref{sec:valid}. Additionally, we are able to make the first comparison of the
data to a theory prediction that includes both NNLO and $\DeltaNNNLO$. We then turn our attention to more recent data taken at $\sqrt{s}=13$~TeV and concentrate on the region
of high invariant mass of the diphoton pair, which is relevant for searches for new physics. This region of phase space is particularly interesting
given the recent observations of excesses in the data at around $750$ GeV~\cite{ATLAS-CONF-2015-081,CMS-PAS-EXO-15-004}. For the remainder of this paper 
we will use the NNLO CT14 PDF set~\cite{Dulat:2015mca} for all predictions (NNLO, NLO, and $\DeltaNNNLO$). The NLO (and $\DeltaNNNLO$) contributions are computed using dipole subtraction
and the NNLO coefficients use jettiness regularization with a value of $\tau^{\rm{cut}} = 0.002$ GeV.  From the studies of section~\ref{sec:valid} we expect this
to give us control of the power corrections at the few per-mille level in the total cross-section. We maintain the EW parameters from the previous section, namely $\alpha = 1/137$. 

\subsection{$pp\rightarrow \gamma\gamma$ as a probe of hard QCD}

As a benchmark we take the recent study by CMS at 7 TeV~\cite{Chatrchyan:2014fsa}.  In order to mimic the cuts applied in the experimental analysis we enforce the following 
phase space selection cuts, 
\begin{eqnarray}
&&p_T^{\gamma,{\rm hard}} > 40 \; {\rm{GeV}}, \quad p_T^{\gamma,{\rm soft}} > 25 \; \rm{GeV} \, , \nonumber\\
&&|\eta_{\gamma} | < 2.5 \quad {\rm{omitting~the~region,}} \; 1.44 < |\eta_{\gamma} | < 1.57 \, , \nonumber\\
&&R_{\gamma\gamma} > 0.45 \, . \nonumber
\end{eqnarray}
Note that the small slice of rapidity that is excluded is due to the design of the CMS detector.
In addition we apply isolation cuts to the photon using the smooth cone prescription~\cite{Frixione:1998jh} that does not require an implementation
of photon fragmentation.  As part of their study CMS compared various smooth cone implementations to that of Diphox, which includes the fragmentation contribution, ultimately employing the following isolation prescription, 
\begin{eqnarray}
E_T^{{\rm{iso}}}(\Delta R)  <  \epsilon \left( \frac{1-\cos {\Delta R}}{1-\cos {R_0}} \right)^n \, , 
\end{eqnarray}
with $\epsilon = 5$ GeV, $R_0 =0.4$ and $n=0.05$. The rather low value of $n$ results in a fairly weak damping of the collinear 
singularity present in the calculation as $\Delta R \rightarrow 0$. Therefore at the cost of deviating from the isolation requirement outlined in 
ref.~\cite{Chatrchyan:2014fsa}, we instead use the following definition, 
\begin{eqnarray}
E_T^{{\rm{iso}}}(\Delta R)  <  \epsilon_{\gamma} p_T^{\gamma} \left( \frac{1-\cos {\Delta R}}{1-\cos {R_0}} \right)^n \, , 
\end{eqnarray}
with $\epsilon_{\gamma} = 0.1$ and $n=2$. We have tuned the values of $\epsilon_\gamma$
and $n$
such that our NLO smooth cone cross section agrees with the theory prediction obtained at NLO with the CMS isolation experimental requirement and GdRG
fragmentation functions~\cite{GehrmannDeRidder:1997gf}.  Choosing such a value of $n$ results in a much more efficient Monte Carlo code. A related study of photon plus jets~\cite{Bern:2011pa} drew similar conclusions. 
We do not believe that the difference in isolation is a particular cause for concern~\cite{Campbell:2014yka}, especially since the cross section has been
tuned to a NLO calculation that includes the effects of fragmentation. In principle, the isolation procedure used by CMS in their theory predictions could be chosen in MCFM, but the calculation of the corresponding NNLO corrections would require significantly more Monte Carlo statistics to evaluate, with little additional benefit. 

We begin by comparing the total cross section as measured by CMS to our prediction using MCFM. The value reported by CMS is, 
\begin{eqnarray}
\sigma^{CMS} = 17.2  \pm 0.2 \; (\rm{stat}) \pm 1.9 \; (syst) \pm 0.4 \; (lumi) \; pb  \, , 
\end{eqnarray}
while our NNLO prediction is 
\begin{eqnarray}
\sigma^{NNLO} = 16.1^{+0.5}_{-0.8} \;(\rm{scale})\; pb \, .
\end{eqnarray}
Thus, within the theoretical and experimental uncertainties, the two are in good agreement. Including the NLO corrections to
the $gg$ initiated pieces raises the theoretical prediction by around 7\%,
\begin{eqnarray}
\sigma^{NNLO}+\DeltaNNNLO = 17.3^{+0.8}_{-0.9} \;(\rm{scale})\; pb \, . 
\end{eqnarray}
Since we do not include the full N$^{3}$LO prediction 
we do not obtain any improvement in the scale variation when including the $gg$ box contributions at NLO. 

\begin{figure}
\begin{center}
\includegraphics[width=0.6\textwidth]{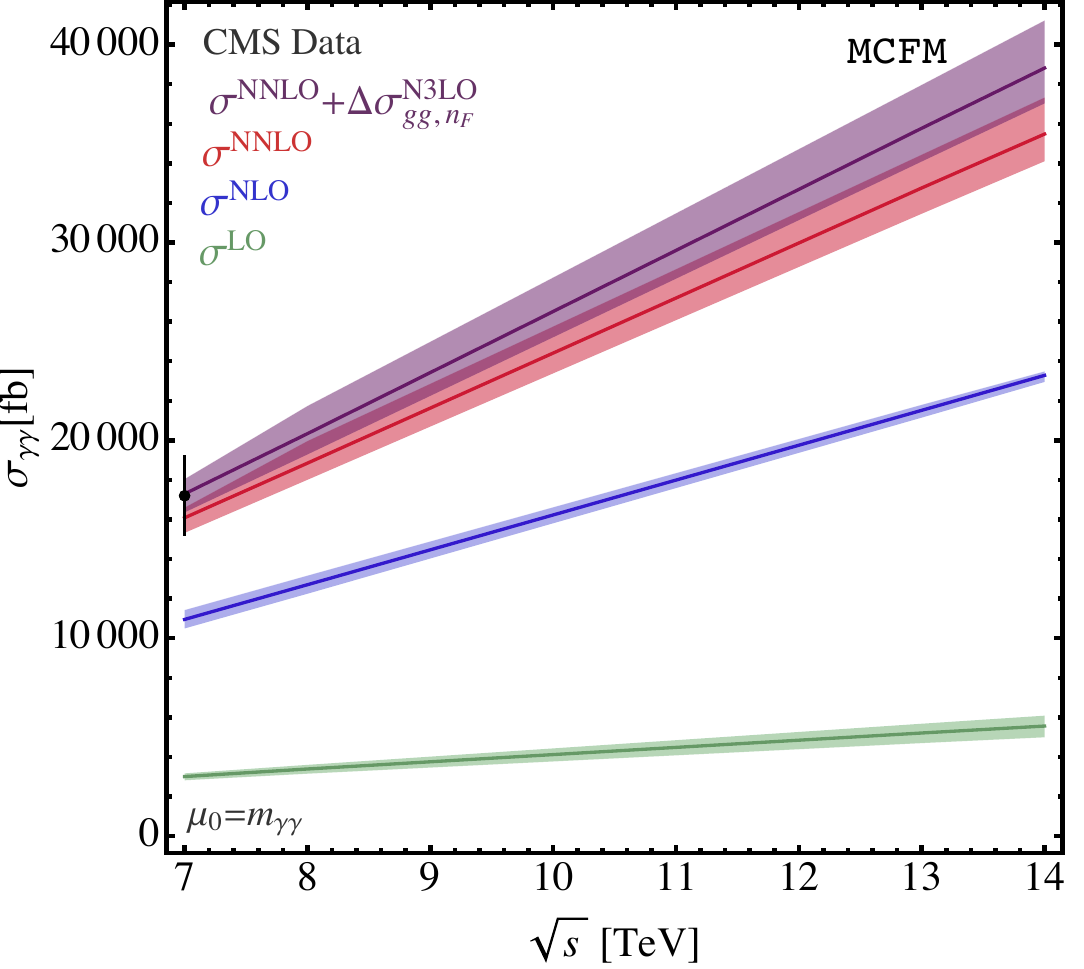}
\caption{The $pp \to \gamma\gamma$ cross section at various orders in perturbation theory, as a function of the LHC operating energy,
$\sqrt{s}$.  Acceptance cuts have been applied, as described in the text.  Also shown is the CMS measurement, under the same set of
cuts, at $7$~TeV~\cite{Chatrchyan:2014fsa}.}
\label{fig:xsCMS}
\end{center}
\end{figure}

As a brief aside, in Figure~\ref{fig:xsCMS} we show the cross section computed at higher center of mass energies, from the
$7$~TeV result discussed above to the highest design energy of the LHC, $14$~TeV.
In the figure we include the cross sections computed at LO, NLO, NNLO and NNLO$+ gg$ boxes at NLO.  As the order in perturbation
theory increases there are sizeable corrections. Going from LO to NLO the cross section increases by around a factor  of 4. The
corrections going from NLO to NNLO are around 1.5. Including the additional $gg$ contributions at NLO increases the cross section by
about a further 10\%. At the 13 TeV LHC the difference between $\sigma^{NNLO}$ and $\sigma^{NNLO} + \DeltaNNNLO$ is more
apparent and it is entirely possible that a measurement will prefer one value over the other. Note that it is not trivially true
that  $\sigma^{NNLO} + \DeltaNNNLO$ is a better prediction than $\sigma^{NNLO}$ since the former is not a complete
N$^3$LO calculation. The missing pieces are not positive definite, and may reduce the cross section such that $\sigma^{N3LO}$ lies
completely within the uncertainty bands of the NNLO calculation. It will be interesting to compare the measured cross sections at $13$~TeV
and $14$~TeV to the two predictions to see if indeed $\sigma^{NNLO} + \DeltaNNNLO$ does a better job of
describing the data than  $\sigma^{NNLO}$ alone.

\begin{figure}
\begin{center}
\includegraphics[width=0.48\textwidth]{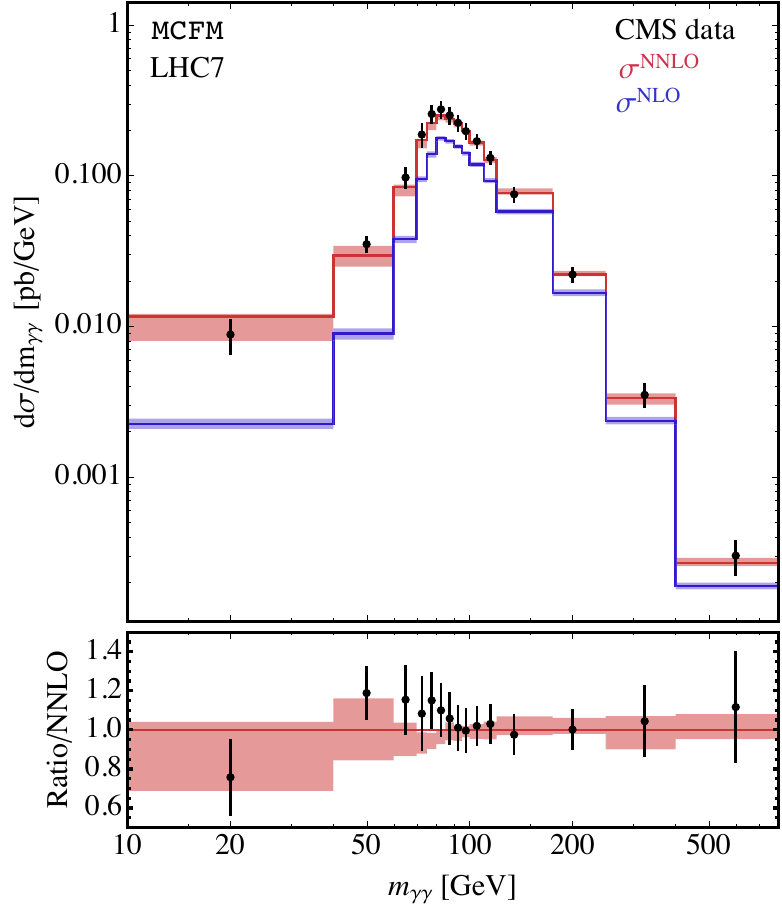}
\includegraphics[width=0.48\textwidth]{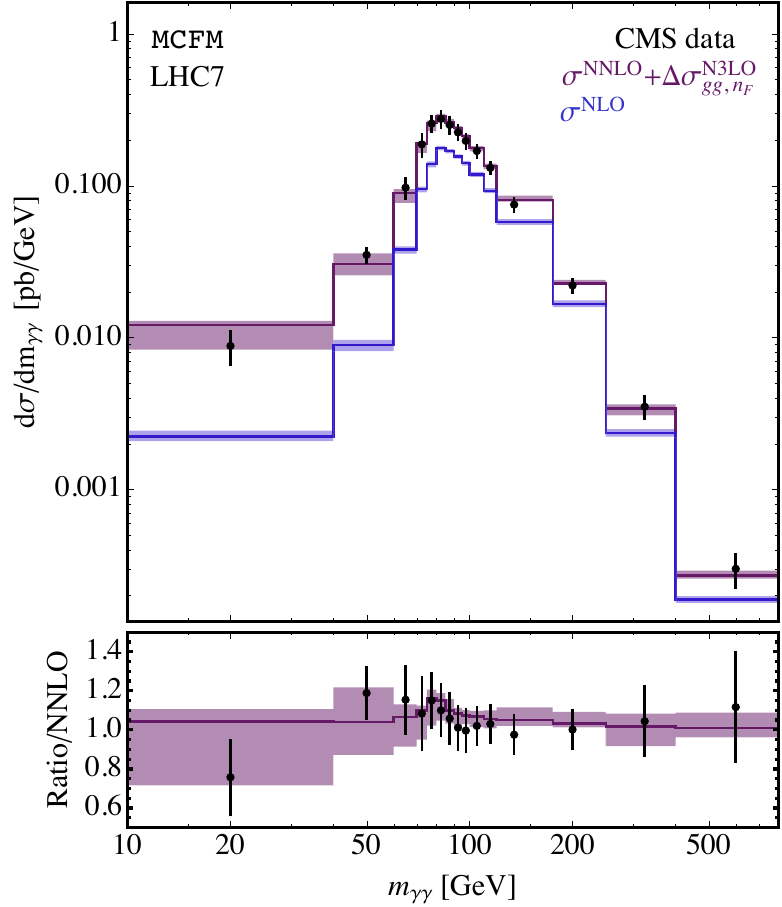}
\caption{The invariant mass of the photon pair $m_{\gamma\gamma}$ at NLO and NNLO, compared with the CMS
data from ref.~\cite{Chatrchyan:2014fsa}.  The pure NNLO prediction is shown in the left panel, while the result that
also includes $gg$ $n_F$ contributions that enter at N$^3$LO is depicted in the right panel.
The lower panels present the ratio of the data and NNLO scale variations to the NNLO theory prediction obtained with the central scale.}
\label{fig:mggCMS7}
\end{center}
\end{figure}

\begin{figure}
\begin{center}
\includegraphics[width=0.48\textwidth]{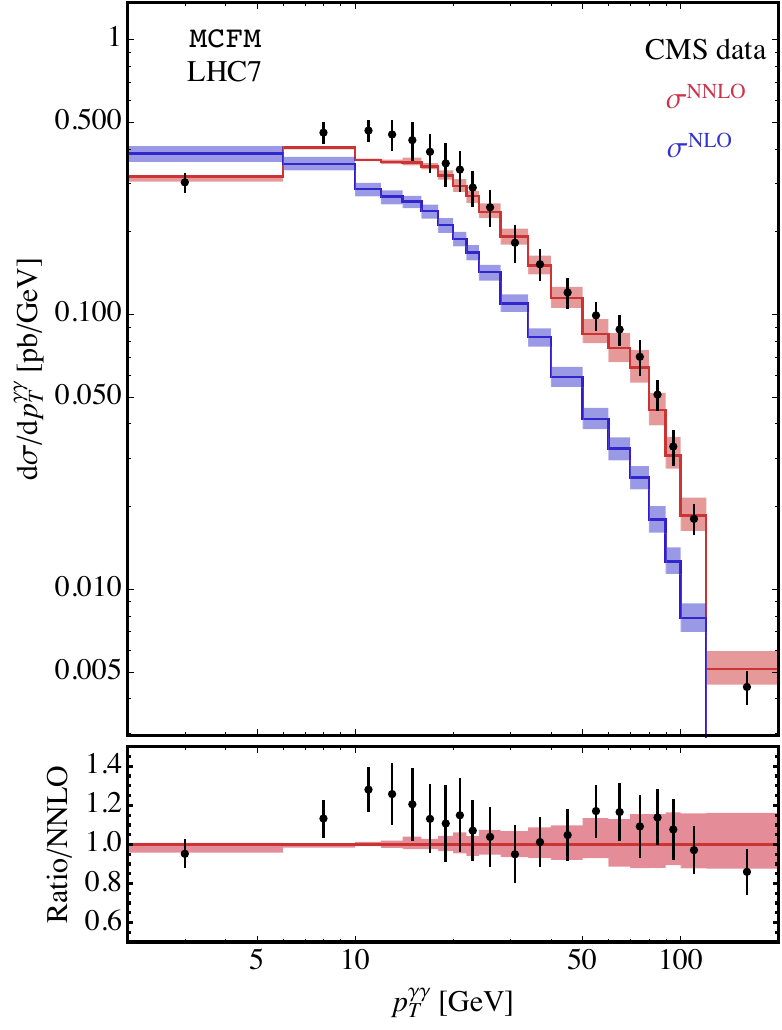}
\includegraphics[width=0.48\textwidth]{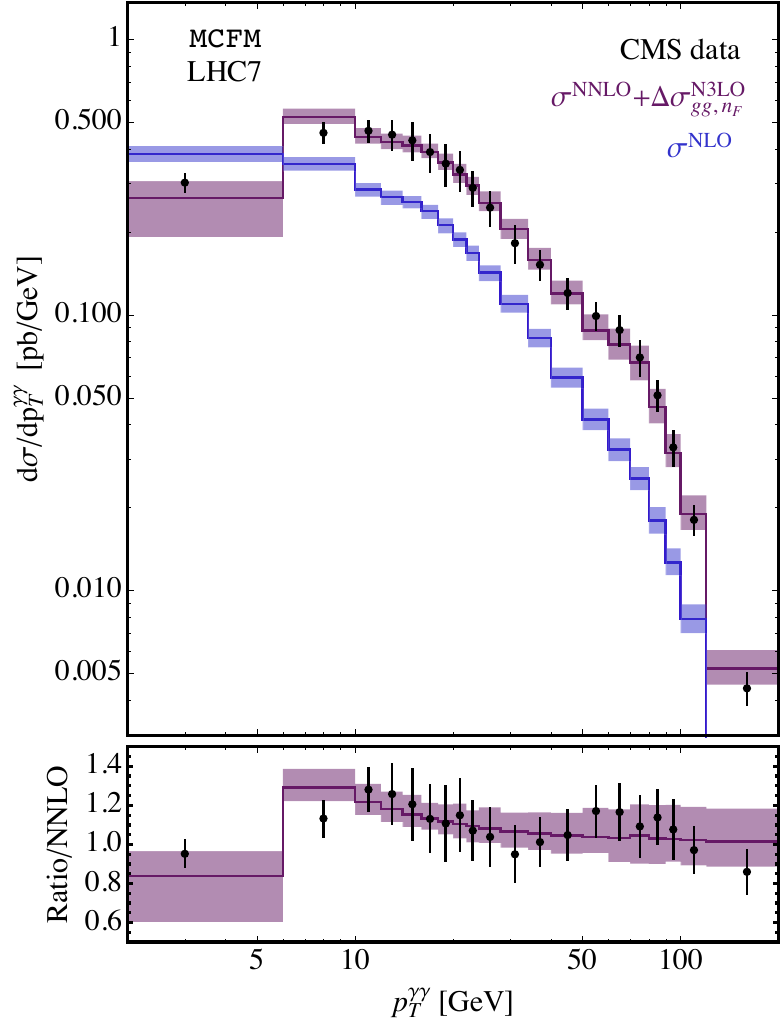}
\caption{As for figure~\ref{fig:mggCMS7}, but for the transverse momentum of the photon pair, $p_T^{\gamma\gamma}$.}
\label{fig:ptggCMS7}
\end{center}
\end{figure}

\begin{figure}
\begin{center}
\includegraphics[width=0.48\textwidth]{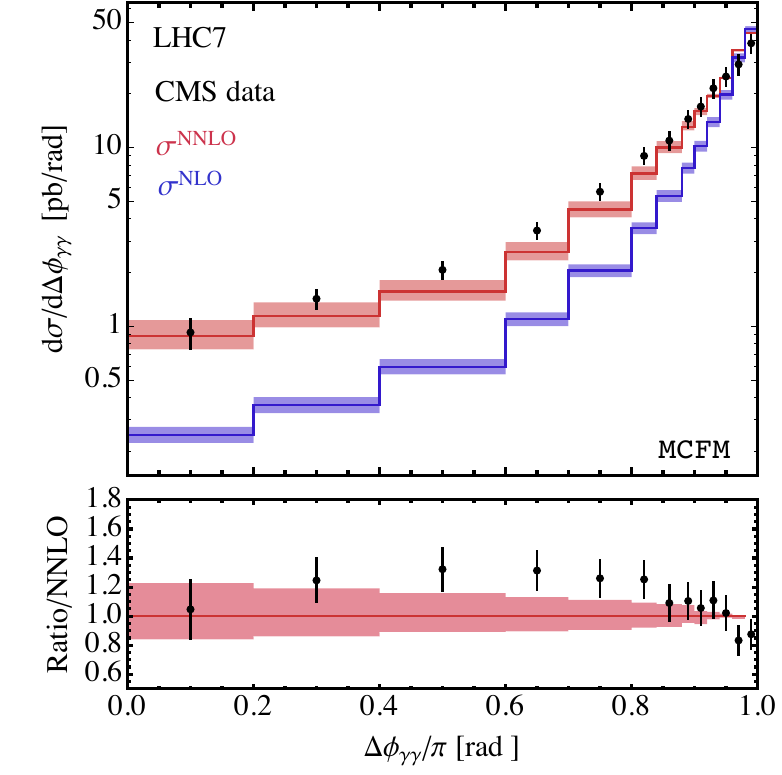}
\includegraphics[width=0.48\textwidth]{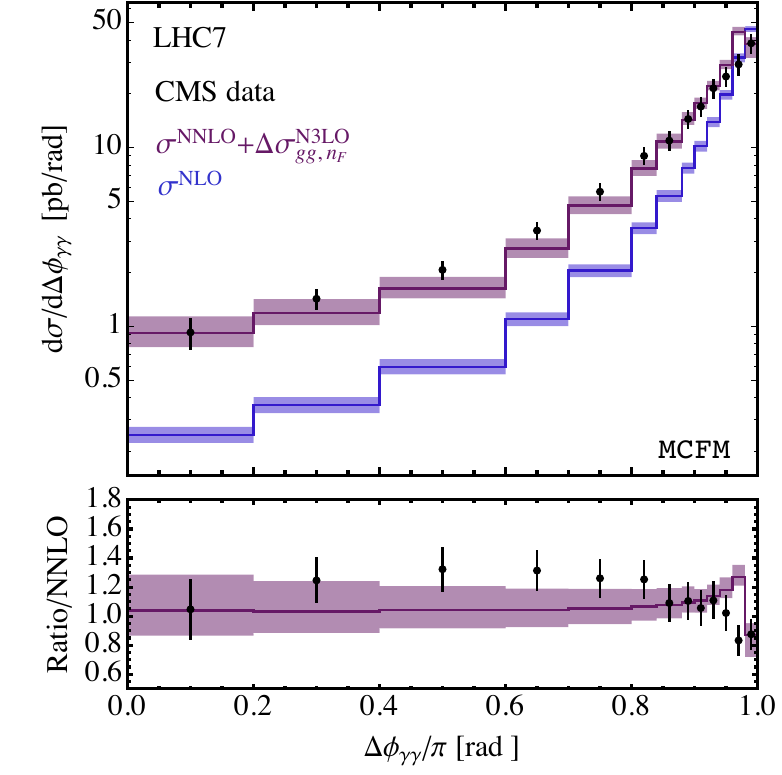}
\caption{As for figure~\ref{fig:mggCMS7}, but for the azimuthal angle between the two photons, $\Delta\phi_{\gamma\gamma}$.}
\label{fig:phiggCMS7}
\end{center}
\end{figure}

We now turn our attention to more differential quantities, namely the invariant mass of the photon pair, $m_{\gamma\gamma}$
(Figure~\ref{fig:mggCMS7}), the transverse momentum of the $\gamma\gamma$ system, $p_T^{\gamma\gamma}$ (Figure~\ref{fig:ptggCMS7}),
and the azimuthal angle between the two photons, $\Delta\phi_{\gamma\gamma}$ (Figure~\ref{fig:phiggCMS7}). We note that, of these
predictions, only $m_{\gamma\gamma}$ is non-trivial at LO since the back-to-back nature of the kinematics at LO means
that $p_T^{\gamma\gamma} = 0$ and $\phi_{\gamma\gamma} = \pi$.
Such distributions that are trivial at LO are particularly sensitive to higher order corrections.
In the bulk of the phase space they first appear at one order higher in $\alpha_s$ 
than the total inclusive cross section. Sadly, most of the distributions made publicly available by the
experimental collaborations suffer from this problem. It would be interesting to additionally compare true NNLO observables, such as the
transverse momenta and rapidities of the photons, in future analyses at higher energies. 

We now examine the predictions for the invariant mass of the photon pair shown in Figure~\ref{fig:mggCMS7} in more detail. Note that
the transverse momentum cuts on the photons requires $m_{\gamma\gamma} >  80$~GeV at LO, so that the region of this distribution
below that value is particularly sensitive to higher order corrections. For all of the figures described here, the plots on the left hand side are
obtained using a pure NNLO prediction, while those on the right represent the prediction obtained with the inclusion of the
$\DeltaNNNLO$ contributions. The NNLO prediction does a good job of describing the data obtained by CMS, although the
central values are typically a little on the low side compared to data. The situation is improved in the right hand plot, after
inclusion of the $\DeltaNNNLO$ pieces. In particular in the region around
$80 \lesssim m_{\gamma\gamma} \lesssim 150$~GeV the prediction follows the shape of the data a little more closely.

In Figure~\ref{fig:ptggCMS7} we turn our attention to the $p_T^{\gamma\gamma}$ spectrum, using the same style as for the
$m_{\gamma\gamma}$ plots. The pure NNLO prediction again describes the data very well, even in the very soft $p_T^{\gamma\gamma} <
10$~GeV region of phase space. Including the $gg$ pieces at NLO improves the agreement with data in the region $
p_T^{\gamma\gamma} > 10$~GeV. In the soft region of phase space it is difficult to argue that the inclusion of the additional
pieces improves the agreement with data. This is understandable since the softest bins are described only after a delicate
cancellation between the various real and virtual pieces of the calculation. By only including a subset of the N$^3$LO calculation we
are unlikely to improve this bin. However in the bulk of the phase space we are typically interested in the types of correction that
are sensitive to the staggered phase space cuts. This is exactly the places where we expect the $gg\rightarrow \gamma\gamma g$
contribution to be important. By including these pieces we therefore do a better job of describing the data.

The situation with the $\Delta\phi_{\gamma\gamma}$ distribution is similar. The NLO prediction for this observable does a very bad
job of describing the CMS data. However by including the NNLO corrections we get much closer to the data, whilst still observing
deviations from the experimental data of order 20\%. Thus, this observable clearly
requires at least a full N$^3$LO prediction to match the experimental data. However, our partial prediction does not do much better.
Again we are exposed to the LO phase space sensitivity in the bins around $\pi$ where it is entirely possible that reasonably large corrections from the three-loop triple virtual and real-double virtual may drive the theoretical prediction down towards the data.

\subsection{Studies of $\gamma\gamma$ at high invariant masses}

One of the most interesting phenomenological aspects of the diphoton production channel during Run II at the LHC is its ability to search 
for new resonances that may manifest themselves in the $m_{\gamma\gamma}$ spectrum. In particular a recent observation of an 
excess around $750$~GeV in the ATLAS experiment~\cite{ATLAS-CONF-2015-081},
with a smaller excess in the same region reported by CMS~\cite{CMS-PAS-EXO-15-004}, has caused considerable 
excitement in the theoretical community. In these analyses the Standard Model background is accounted for by using a data-driven approach
that fits a smooth polynomial function to the data across the entire $m_{\gamma\gamma}$ spectrum.  A resonance might then be observed
as a local excess in this spectrum, deviating from the fitted form.  Although well-motivated, one might be concerned that the
spectrum may not be correctly modeled at high energies, where there is little data, and that small fluctuations could unduly influence
the form of the fit and result in misinterpretation of the data.  Such worries could be lessened by using a first-principles 
theoretical prediction for the spectrum and it is this issue that we aim to address in this section.

As a concrete example, we will produce NNLO predictions for the invariant mass spectrum at high energies using cuts that are inspired
by the recent ATLAS analysis~\cite{ATLAS-CONF-2015-081}. Specifically, these are:
\begin{eqnarray}
&&p_T^{\gamma,{\rm hard}} > 0.4 \, m_{\gamma\gamma},  \quad  p_T^{\gamma,{\rm soft}} > 0.3 \, m_{\gamma\gamma},  \nonumber \\
&& |\eta^{\gamma} | < 2.37,  \quad {\rm{excluding~the~region,}} \; 1.37 < |\eta_{\gamma} | < 1.52.  
\end{eqnarray}
We will only be interested in the region $m_{\gamma\gamma} > 150$~GeV, so these represent hard cuts on the photon momenta.
The small region of rapidity that is removed corresponds to the transition from barrel to end-cap calorimeters.
We maintain the same isolation requirements as the previous section, which again differs slightly from the treatment in the
ATLAS paper.

\begin{figure}
\begin{center}
\includegraphics[width=0.75\textwidth]{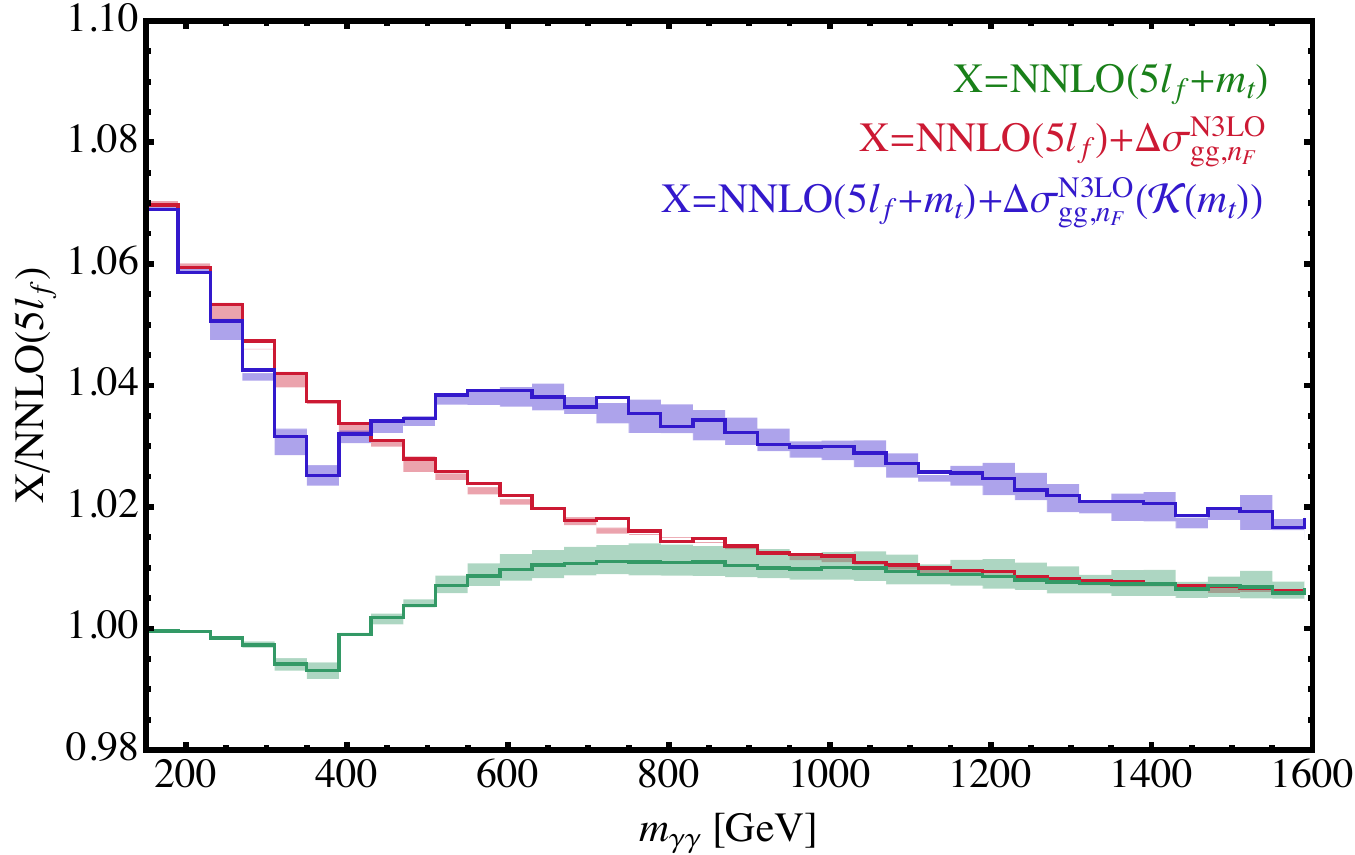}
\caption{The ratio of various different theoretical predictions to the NNLO $n_F=5$ differential cross section.
The different predictions correspond to: the inclusion of the  top quark $gg\rightarrow \gamma\gamma$ box diagrams (green),
the $\DeltaNNNLO$ correction (red) and the $\DeltaNNNLO$ and the top boxes with the
$\DeltaNNNLO$ correction re-scaled by the ratio $\mathcal{K}(m_t)$ described in the text (blue).}
\label{fig:mggratios}
\end{center}
\end{figure} 
Our first concern is to address the impact of the $gg$ pieces at NLO, represented by the contribution $\DeltaNNNLO$ defined previously, and the
contribution of the top quark loop. We summarize our results in Figure~\ref{fig:mggratios}, in which we present several different theoretical
predictions, each normalized to the the default NNLO prediction with 5 light flavors.  The first alternative is  one in which the NNLO
prediction is augmented by the inclusion of the top loops, i.e. the $gg$ contribution corresponds to $\sigma_{gg}(m_t+5l_f)$ in the
notation of section~\ref{sec:toploops}. In the second prediction we use the result for five light flavors but add the NLO corrections
to the $gg$ channel, i.e. the term  $\DeltaNNNLO$.  For the final alternative we include the top quark loop contribution
and attempt to account for the NLO corrections to all $gg$ loops by rescaling the $\DeltaNNNLO$ result by
a factor $\mathcal{K}(m_t)$ that is given by,
\begin{eqnarray}
\mathcal{K}(m_t) = \frac{ \sigma_{gg}(5\ell_f + m_t)}{ \sigma_{gg}(5\ell_f )} \, .
\end{eqnarray}
This collection of predictions covers a range of theoretical options that may extend the NNLO predictions described in the previous sections. The
top loops, illustrated by the green curve in the figure typically represent around a $1\%$ effect across the invariant mass range of interest.
For $m_{\gamma\gamma} < 2m_t$ there is a destructive interference, which reduces the cross section, whilst at higher energies there is a small
enhancement. Therefore, although the top loops are an important contribution in terms of the $n_F$ box loops (as shown in section~\ref{sec:calc}),
they are not particularly important in the total rate. 
At this order the $gg$ pieces reside in the Born phase space, which is particularly impacted by the staggered cuts at high $m_{\gamma\gamma}$.

 As we found in the previous section  the effects of the NLO corrections to the $gg$ pieces are larger, however their effects are much
more pronounced at lower invariant masses. By the time invariant masses of order $500$~GeV are probed, the corrections are $2\%$ or smaller. The
attempt to model the combined effect of corrections to both the light-quark and top quark loops shows, as expected, the largest deviations from
the NNLO($5\ell_f$) prediction. However the deviations are still of order $3\%$ or smaller in the high invariant mass region.
Therefore, although the corrections to the $gg$ loops and the effect of the finite top quark mass can have about a 6\%
effect at invariant masses around $200$~GeV, the effect at higher masses is somewhat smaller. Since we aim to compare the ATLAS data, which is
not corrected for fakes or identification efficiencies, to our parton-level prediction we are not concerned about effects at this level.  
As a result we will simply use the most consistent prediction\footnote{This is because a consistent inclusion of the effect of top quark loops
would require alterations to the running of $\alpha_s$ and additional top quark loops in the $q\overline{q}g\gamma\gamma$ one-loop amplitude.}, corresponding to NNLO($5\ell_f$), for comparison with the fitting
function used by ATLAS.

\begin{figure}
\begin{center}
\includegraphics[width=0.6\textwidth]{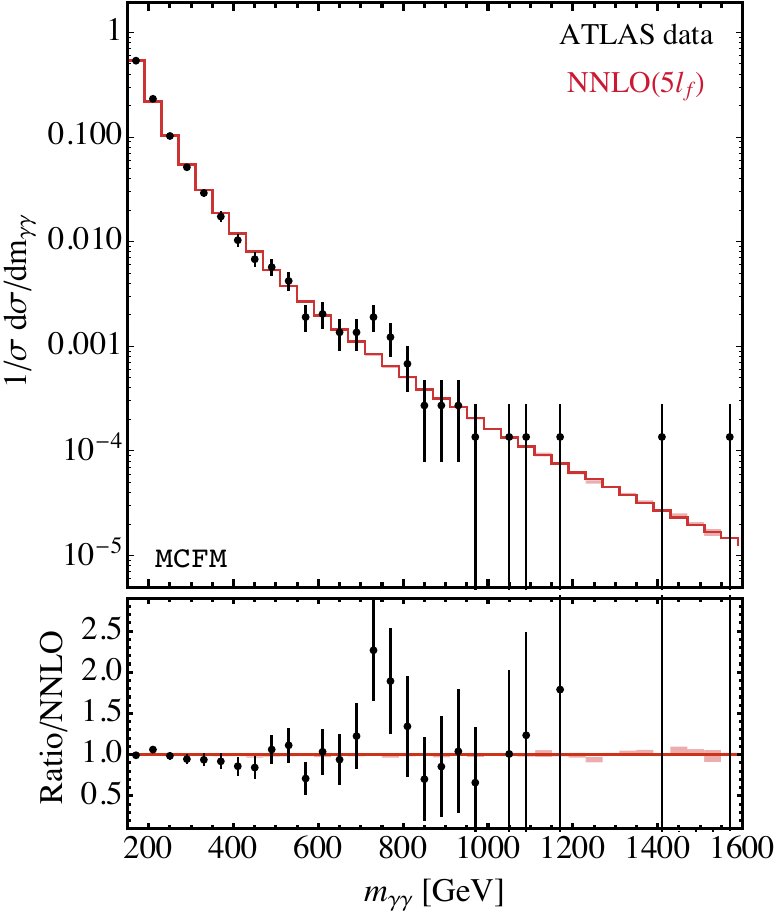}
\caption{The rate-normalized shapes of the $m_{\gamma\gamma}$ distribution from
the ATLAS collaboration and the MCFM NNLO prediction for $\mu=m_{\gamma\gamma}$.
The lower panel indicates the ratio of the data to the NNLO prediction.}
\label{fig:mggATLAS}
\end{center}
\end{figure} 
We compare our NNLO prediction to the ATLAS data in Figure~\ref{fig:mggATLAS}. We note that to properly compare our prediction to the data
requires knowledge of both the fake rate and the photon efficiencies and acceptance corrections of the ATLAS detector. To try to minimize the
impact of such corrections we simply compare the shape of the ATLAS data to the shape of our NNLO prediction, i.e. we normalize our prediction to
$1/\sigma^{NNLO}$ and the ATLAS data to $1/N_{\rm{events}}$. From this comparison we can draw several conclusions. First, we note that our
prediction is in excellent agreement with the overall shape of the data, indicating that the theoretical prediction for the shape of the
$m_{\gamma\gamma}$ distribution could easily be used in place of the somewhat arbitrary fitting functions currently employed. Second, the
excellent agreement in shape suggests either a low number of fakes, or that the fake events are distributed with a similar shape to the
Standard Model prediction for the $\gamma\gamma$ spectrum.   Of course  a combination of these two explanations is also possible.
Finally, and most excitingly, a comparison
to the fitting function presented in ref.~\cite{ATLAS-CONF-2015-081} illustrates that there is no significant hardening from the prediction of
the SM compared to the form of the fitting function used in the ATLAS experiment. This can clearly be seen upon comparison with Figure~1 in
ref.~\cite{ATLAS-CONF-2015-081}.  
For instance, both the ATLAS fit and our NNLO prediction pass directly through the data in the 1090 GeV bin, and just under the central value in the 690 GeV bin. 
Therefore we can conclude that the interpretation of an excess
of events around $750$~GeV appears to be supported by a first-principle calculation within the SM.  It is not diluted by a hardening of the SM spectrum relative to the
fitting function used in the analysis. If the excess is confirmed, NNLO predictions for the shape of the irreducible background will be able to
significantly enhance analyses designed to discriminate between different model hypotheses, by providing predictions for the
properties of background events that cannot be captured by a simple spectrum fit.

\section{Conclusions} 
\label{sec:conc}

The process $pp\rightarrow \gamma\gamma$ is a flagship process for Run II phenomenology.  Besides its intrinsic interest as
a tool to understand the perturbative nature of QCD at high energies, it represents an important background in studies
of the Higgs boson that are  a cornerstone of the Run II physics program.  In addition it is a clean and well-measured final state
that can be used in the search for new heavy resonances.   Often these analyses require staggered photon transverse momentum cuts
that induce large corrections at higher orders in perturbation theory. Essentially the NLO prediction behaves like a LO prediction
since the staggered cuts are first accessible at this order.  This therefore necessitates the inclusion of NNLO corrections to capture
the corrections to the rate in this larger phase space and hence adequately describe data.  

In this paper we have presented a NNLO calculation of the process $pp\rightarrow \gamma\gamma$ and studied the phenomenology of this process at
the LHC.  We have used the recently-developed $N$-jettiness slicing procedure to manage the infrared singularities present in the NNLO
calculation and have implemented the calculation into the  Monte Carlo MCFM.  The calculation will be made available in a forthcoming release of
the code.  Given the signficant effect of the NNLO corrections to this process, our slicing procedure is subject to large power corrections
and care must be taken to ensure that a small enough value of the slicing parameter is employed.
We have compared our results to an existing calculation of the same process and found that we could not reproduce the results present
in the literature, despite extensive testing and investigation.
However we have communicated with the authors of ref.~\cite{Catani:2011qz} and believe that, after correction of a bug in their numerical code,
their results will be consistent with ours.

We have used our calculation to compare to data obtained at $7$~TeV by CMS and to $13$~TeV data collected by the ATLAS experiment. The latter is
particularly exciting  given the excess reported in the data at around $m_{\gamma\gamma} \sim 750$ GeV. We found that the shape of our NNLO
prediction does a good job  of describing the experimental data, and simultaneously has a good agreement with the fitting function used by the
ATLAS collaboration. We therefore do  not expect further data at high energies to dramatically alter the form of the fit used by the
collaboration.  Furthermore, we do not believe that the excess is due to the use of a fitting function that underestimates the prediction of the
SM at high invariant masses. 

\acknowledgments
We thank Felix Yu for providing us with an extraction of the ATLAS diphoton invariant mass spectrum. 
JC is grateful to the Fermilab computing sector for
providing access to the Accelerator Simulations Cluster.
Fermilab is supported by the US DOE under contract DE-AC02-07CH11359. 
Support provided by the Center for Computational Research at the University at Buffalo.

\appendix

\section{Ingredients for $q\overline{q} \rightarrow \gamma\gamma$ at NNLO}
\label{app:cal}

\subsection{Below $\tau^{\rm{cut}}$: Hard function} 

The virtual matrix elements needed to compute $q\overline{q}\rightarrow \gamma\gamma$ to  $\mathcal{O}(\alpha_s^2)$ accuracy can be found in ref.~\cite{Anastasiou:2002zn}. 
In order to be utilized in $N$-jettiness slicing these results must be translated into the form of a SCET hard function. This can be 
achieved using the procedure outlined, for instance, in refs.~\cite{Becher:2009qa,Becher:2009cu}. We begin by defining the UV-renormalized
matrix element as follows,
\begin{eqnarray}
|\mathcal{M}_{q\overline{q}\gamma\gamma}\rangle= 4\pi \alpha \left[|\mathcal{M}^{(0)}_{q\overline{q}\gamma\gamma}\rangle
+\left(\frac{\alpha_s}{2\pi}\right)|\mathcal{M}^{(1)}_{q\overline{q}\gamma\gamma}\rangle
+\left(\frac{\alpha_s}{2\pi}\right)^2|\mathcal{M}^{(2)}_{q\overline{q}\gamma\gamma}\rangle+\mathcal{O}(\alpha_s^3)\right] \, ,
\end{eqnarray}
where $\alpha_s$ is the renormalized strong coupling, and $\alpha$ is the (bare) electromagnetic coupling.
The matrix elements are defined in terms of Mandelstam invariants $s=(p_1+p_2)^2$, $t=(p_1+p_3)^2$ and $u=(p_1+p_4)^2$, with $s+t+u=0$.
For the process under investigation $p(-p_1)+p(-p_2)\rightarrow \gamma(p_3)+\gamma(p_4)$ we have $s > 0$ while $t$,~$u < 0$.
Following the notation of ref.~\cite{Anastasiou:2002zn} we define the matrix element squared as follows, 
\begin{eqnarray}
\sum |\mathcal{M}(q(-p_1)+\overline{q}(-p_2)\rightarrow \gamma(p_3)+\gamma(p_4))|^2 = \mathcal{A}_{q\overline{q}\gamma\gamma}(s,t,u) \, .
\end{eqnarray}
Expanding to $\mathcal{O}(\alpha_s^2)$ we define,
\begin{eqnarray}
\mathcal{A}_{q\overline{q}\gamma\gamma}(s,t,u) &=& 16\pi^2\alpha^2 \bigg[
\mathcal{A}^{\rm{LO}}_{q\overline{q}\gamma\gamma}(s,t,u)  +
\left(\frac{\alpha_s}{2\pi}\right)\mathcal{A}^{\rm{NLO}}_{q\overline{q}\gamma\gamma}(s,t,u)  + \nonumber\\&& 
\left(\frac{\alpha_s}{2\pi}\right)^2\mathcal{A}^{\rm{NNLO}}_{q\overline{q}\gamma\gamma}(s,t,u) +\mathcal{O}(\alpha_s^3) \bigg] \, .
\end{eqnarray}
In terms of the matrix elements defined above we have 
\begin{eqnarray}
\mathcal{A}^{\rm{LO}}_{q\overline{q}\gamma\gamma}(s,t,u) &=& \langle \mathcal{M}^{(0)}_{q\overline{q}\gamma\gamma} |  \mathcal{M}^{(0)}_{q\overline{q}\gamma\gamma} \rangle \, ,  \\
\mathcal{A}^{\rm{NLO}}_{q\overline{q}\gamma\gamma}(s,t,u) &=& \langle \mathcal{M}^{(0)}_{q\overline{q}\gamma\gamma} |  \mathcal{M}^{(1)}_{q\overline{q}\gamma\gamma} \rangle + \langle \mathcal{M}^{(1)}_{q\overline{q}\gamma\gamma} |  \mathcal{M}^{(0)}_{q\overline{q}\gamma\gamma} \rangle \, ,  \\
\mathcal{A}^{\rm{NNLO}}_{q\overline{q}\gamma\gamma}(s,t,u) &=& \mathcal{A}^{\rm{NNLO(0\times2)}}_{q\overline{q}\gamma\gamma}(s,t,u) +\mathcal{A}^{\rm{NNLO(1\times1)}}_{q\overline{q}\gamma\gamma}(s,t,u) \, .
\end{eqnarray}
where
\begin{eqnarray}
 \mathcal{A}^{\rm{NNLO(0\times2)}}_{q\overline{q}\gamma\gamma}(s,t,u)  &=& \langle \mathcal{M}^{(0)}_{q\overline{q}\gamma\gamma} |  \mathcal{M}^{(2)}_{q\overline{q}\gamma\gamma} \rangle +  \langle \mathcal{M}^{(2)}_{q\overline{q}\gamma\gamma} |  \mathcal{M}^{(0)}_{q\overline{q}\gamma\gamma} \rangle  \, , \\
  \mathcal{A}^{\rm{NNLO(1\times1)}}_{q\overline{q}\gamma\gamma}(s,t,u)  &=&  \langle \mathcal{M}^{(1)}_{q\overline{q}\gamma\gamma} |  \mathcal{M}^{(1)}_{q\overline{q}\gamma\gamma} \rangle \, .
\end{eqnarray}
The aim of this section is to re-write the above expressions in the SCET renormalized form, which is obtained via the following re-definitions~\cite{Becher:2009qa,Becher:2009cu}
\begin{eqnarray}
|\mathcal{M}^{(1),\text{ren}}_{q\overline{q}\gamma\gamma}\rangle&= &
|\mathcal{M}^{(1),\text{fin}}_{q\overline{q}\gamma\gamma}\rangle+ \left(\bI^{(1)}(\epsilon) +\boldsymbol{Z}^{(1)}(\epsilon)\right) |\mathcal{M}^{(0)}_{q\overline{q}\gamma\gamma}\rangle \, , \\
|\mathcal{M}^{(2),\text{ren}}_{q\overline{q}\gamma\gamma}\rangle&=&
|\mathcal{M}^{(2),\text{fin}}_{q\overline{q}\gamma\gamma}\rangle+
\left(\bI^{(1)}(\epsilon) +\boldsymbol{Z}^{(1)}(\epsilon)\right)
 |\mathcal{M}^{(1),\text{fin}}_{q\overline{q}\gamma\gamma}\rangle  \nonumber\\&&
 + \Big(\bI^{(2)}(\epsilon) + \left(\bI^{(1)}(\epsilon) +\boldsymbol{Z}^{(1)}(\epsilon)\right) \bI^{(1)}(\epsilon) +
\boldsymbol{Z}^{(2)}(\epsilon)\Big) |\mathcal{M}^{(0)}_{q\overline{q}\gamma\gamma}\rangle \, .
\end{eqnarray}
$\bI^{(1)}(\epsilon)$ and $\bI^{(2)}(\epsilon)$ are obtained via Catani's IR-subtraction formula~\cite{Catani:1998bh}. For the $q\overline{q}\gamma\gamma$ process under investigation here $\bI^{(1)}(\epsilon)$ and $\bI^{(2)}(\epsilon)$ are defined as follows
\begin{eqnarray}
\bI^{(1)}(\epsilon)&=&-C_F\frac{e^{\epsilon \gamma_E}}{\Gamma(1-\epsilon)}\left(\frac{1}{\epsilon^2}+\frac{3}{2\epsilon}\right)\left(\frac{\mu^2}{-s}\right)^{\epsilon} \\
\bI^{(2)}(\epsilon)&=& -\frac{1}{2}\bI^{(1)}(\epsilon)\left(\bI^{(1)}(\epsilon)+\frac{\beta_0}{\epsilon}\right)\nonumber\\&& +\frac{e^{-\epsilon \gamma_E}\Gamma(1-2\epsilon)}{\Gamma(1-\epsilon)}\left(\frac{\gamma^{\rm{cusp}}_1}{8} + \frac{\beta_0}{2\epsilon}\right)\bI^{(1)}(2\epsilon)+H^{2}_{R.S.}(\epsilon) \, .
\end{eqnarray} 
In the above equation the $H^{2}_{R.S.}(\epsilon)$ is a scheme dependent function, containing $1/\epsilon$ poles that
for this process is defined as~\cite{Catani:1998bh}\footnote{While not fully specified for general process in ref.~\cite{Catani:1998bh}, an all-orders form was derived in refs.~\cite{Becher:2009qa,Becher:2009cu}}
\begin{eqnarray}
H^{2}_{R.S.}(\epsilon) =\frac{1}{8 \epsilon}\left(\gamma^q_1 -\frac{\gamma_1^{\rm{cusp}}}{4}\gamma_0^q  +\frac{\pi^2}{16} \beta_0 \gamma_0^{\rm{cusp}}C_F\right).
\end{eqnarray}
$H^{2}_{R.S.}(\epsilon)$ is thus defined in terms of the coefficients of the cusp anomalous dimension $\gamma^{\rm{cusp}}$, quark field
anomalous dimension $\gamma^q$, and the $\beta$ function, that are are given by,
\begin{eqnarray}
&&\gamma_0^{{\rm{cusp}}}  = 4,\nonumber \\
&&\gamma_1^{{\rm{cusp}}}  =  \bigg(\frac{268}{9}-\frac{4\pi^2}{3}\bigg)C_A  - \frac{80}{9}T_f n_f, \\
&&\gamma_0^{q}  = -3C_F, \nonumber \\
&&\gamma_1^{q}  =  \bigg(-\frac{3}{2}+2\pi^2-24\zeta_3\bigg)C_F^2  +C_FC_A\bigg(-\frac{961}{54}-\frac{11\pi^2}{6}+26\zeta_3\bigg)
+C_FT_fn_f\bigg(\frac{130}{27}+\frac{2\pi^2}{3}\bigg).\nonumber
\end{eqnarray}
and
\begin{eqnarray}
\beta_0 = \frac{11}{3}C_A -\frac{4}{3}T_f n_f,
\end{eqnarray}
$\boldsymbol{Z}$ is defined, for our process and order in perturbation theory as~\cite{Becher:2009qa,Becher:2009cu}
\begin{eqnarray}
\boldsymbol{Z}^{(1)}(\epsilon) &=& - \frac{\Gamma'_0 }{8\epsilon^2} - \frac{\boldsymbol{\Gamma}_0}{4\epsilon}, \\
\boldsymbol{Z}^{(2)}(\epsilon) &=& \frac{(\Gamma'_0)^2}{128\epsilon^4}+\frac{3\beta_0\Gamma'_0+2\Gamma'_0\boldsymbol{\Gamma}_0}{64\epsilon^3}+
\frac{4\beta_0\boldsymbol{\Gamma}_0+2\boldsymbol{\Gamma}_0^2-\Gamma'_1}{64\epsilon^2}
-\frac{\boldsymbol{\Gamma}_1}{16 \epsilon}.
\end{eqnarray}
where 
\begin{eqnarray}
\Gamma'_0 &=& -\gamma_0^{{\rm{cusp}}} (2 C_F), \\
\Gamma'_1 &=& -\gamma_1^{{\rm{cusp}}} (2 C_F), \\
\boldsymbol{\Gamma}_0&=&-C_F\gamma_0^{{\rm{cusp}}}\log{\left(\frac{\mu^2}{-s}\right)} + 2\gamma_0^q, \\
\boldsymbol{\Gamma}_1&=&-C_F\gamma_1^{{\rm{cusp}}}\log{\left(\frac{\mu^2}{-s}\right)} + 2\gamma_1^q .
\end{eqnarray}
We can then define our hard functions in terms of the renormalized matrix elements as follows, 
\begin{eqnarray}
\mathcal{\tilde{A}}^{X}_{q\overline{q}\gamma\gamma} = \mathcal{A}^{X}_{q\overline{q}\gamma\gamma} (\mathcal{M}^{(i)}_{q\overline{q}\gamma\gamma} \rightarrow \mathcal{M}^{(i),\text{ren}}_{q\overline{q}\gamma\gamma}).
\end{eqnarray}
For brevity we present the results obtained at $\mu^2=s$;
the full scale dependence may be obtained by inspection of the distributed MCFM routines, or analytically by appropriate usage of the renormalization group equations.
The hard function for the NLO process is given by 
\begin{eqnarray}
 && \mathcal{\tilde{A}}^{\rm{NLO}}_{q\overline{q}\gamma\gamma}(s,t,u,\mu^2=s) =
   \frac{4 C_F}{3 t u} \bigg(12\, t u\, (X + Y+ X^2 + Y^2) \\
 && \qquad + u^2 \left(7 \pi^2 - 6 (7 - 3 X - X^2 - 2 Y^2)\right) + 
   t^2 \left(7 \pi^2 - 6 (7  - 3 Y - 2 X^2 - Y^2)\right)\bigg) \, , \nonumber 
\end{eqnarray}
where we have introduced the following notation~\cite{Anastasiou:2002zn}
\begin{equation}
 X=\log{\left(-\frac{t}{s}\right)},  \quad Y=\log{\left(-\frac{u}{s}\right)} \, , 
\end{equation}
and at NNLO 
\begin{eqnarray}
&& \mathcal{\tilde{A}}^{\rm{NNLO}}_{q\overline{q}\gamma\gamma}(s,t,u,\mu^2=s) =
\mathcal{F}_{inite}^{{1\times1}}(\mu^2=s)+\mathcal{F}_{inite}^{{2\times0}}(\mu^2=s) \nonumber\\&& \quad
-C_AC_F \frac{(t^2+u^2)}{54 tu}\left(-2764\pi^2+75 \pi^4+396 \zeta_3\right) 
-C_FN_F \frac{4(t^2+u^2)}{27 tu}\left(56\pi^2 - 9 \zeta_3\right)  \nonumber\\&& \quad 
+C_F^2 \frac{7\pi^2}{9}\bigg(24(X+X^2+Y+Y^2) + \frac{t}{u}(7 \pi^2 + 12 (-7 + 2 X^2 + 3 Y + Y^2))\nonumber\\&& \qquad
+ \frac{u}{t} (7 \pi^2 + 12 (-7 + 3 X + X^2 + 2 Y^2))\bigg) \, .
\end{eqnarray}
The functions $\mathcal{F}_{inite}^{{1\times1}}$ and $\mathcal{F}_{inite}^{{2\times0}}$ are defined in Eq.~(5.3) and
Eq.~(4.6) of  ref.~\cite{Anastasiou:2002zn}. 
We have adjusted the results of ref.~\cite{Anastasiou:2002zn} to account for a number of small typos in the manuscript, two of which were also noted in ref.~\cite{Catani:2013tia}.
Firstly, we have altered the factor $\Gamma(1-\epsilon)/\Gamma(1-2\epsilon)$ in their Eq. (3.13) to  $\Gamma(1-2\epsilon)/\Gamma(1-\epsilon)$. 
Secondly, the overall sign in equations (C.1), (C.2) and (C.3) must be flipped. Finally, the dressing of the electroweak charges in their Eq.~(4.6) 
is ambiguous. As written the whole of their Eq.~(4.6) is multiplied by the charge of the quark present in the LO matrix element, $Q_j^4$. However 
the first term in the equation, which is associated with a closed loop of fermions, should only be dressed with a factor of $Q_j^2$.
This point is not made explicitly clear in ref.~\cite{Anastasiou:2002zn} but is easily corrected. 

\subsection{Above $\tau^{\rm{cut}}$ } 

For $\tau > \tau^{cut}$ the calculation corresponds to an NLO calculation of the $\gamma\gamma j$ process. An implementation
of this process and the $\gamma\gamma\gamma$ process in the MCFM framework was presented in ref.~\cite{Campbell:2014yka}. We
use the results of this calculation, which corresponds to an analytic calculation using helicity amplitudes and
$D$-dimensional unitarity methods to obtain our above-$\tau^{\rm{cut}}$ pieces. We refer the interested reader to
ref.~\cite{Campbell:2014yka} for more details. 

\section{$gg\rightarrow \gamma\gamma$: $m_t$ loops}
\label{app:top}

In this section we present the calculation of the $gg\rightarrow \gamma\gamma$ process that proceeds through a top-quark loop. 
We use the spinor helicity formalism to define our amplitudes, and refer readers unfamiliar with the notation and conventions to one of the 
many comprehensive reviews of the topic (for instance ref.~\cite{Dixon:1996wi}). Throughout our calculation of these amplitudes we made frequent 
use of the Mathematica package S@M~\cite{Maitre:2007jq}.

We define the partial amplitude for this process as follows, 
\begin{eqnarray}
A^{(1),m_t}_4(1_g^{h_1},2_g^{h_2},3_\gamma^{h_3},4_\gamma^{h_4}) = 2 Q_t^2 e^2 \alpha_s^2 \delta^{a_1a_2}\mathcal{A}^{(1),m_t}_4(1_g^{h_1},2_g^{h_2},3_\gamma^{h_3},4_\gamma^{h_4} ) \, .
\end{eqnarray}
The simplest amplitude corresponds to the case where all of the bosons have the same helicity 
\begin{eqnarray}
\mathcal{A}^{(1),m_t}_4(1_g^{+},2_g^{+},3_\gamma^{+},4_\gamma^{+} )&=&2\frac{\spb 1.2 \spb3.4}{\spa3.4 \spa 1.2}\bigg(\frac{1}{2}-m_t^4 \big(I_4(s_{13},s_{12},m_t^2)\nonumber\\&&+I_4(s_{14},s_{12},m_t^2)+I_4(s_{13},s_{14},m_t^2)\big)\bigg) \, .
\end{eqnarray}
The next simplest case corresponds to either a single photon or single gluon having negative helicity. 
\begin{eqnarray}
\mathcal{A}^{(1),m_t}_4(1_g^{+},2_g^{-},3_\gamma^{+},4_\gamma^{+} )&=&
m_t^2\left(\frac{\spa1.2^2\spb3.1^2}{\spa1.4^2}- 2m_t^2 \frac{\spb3.1\spb4.1\spb4.3}{\spa3.4\spb3.2\spb4.2}\right)I_4(s_{13},s_{12},m_t^2)\nonumber\\&&
+m_t^2\left(\frac{\spa1.2^2\spb4.1^2}{\spa1.3^2}- 2m_t^2 \frac{\spb3.1\spb4.1\spb4.3}{\spa3.4\spb3.2\spb4.2}\right)I_4(s_{14},s_{12},m_t^2)\nonumber\\&&
+m_t^2\left(\frac{\spb3.1^2\spb4.1^2}{\spb2.1^2}- 2m_t^2 \frac{\spb3.1\spb4.1\spb4.3}{\spa3.4\spb3.2\spb4.2}\right)I_4(s_{13},s_{14},m_t^2)\nonumber\\
&&+2m_t^2 \frac{\spb3.1}{\spb2.1}\left(\frac{\spa1.2\spb3.1}{\spa2.4^2}-\frac{\spa2.3\spb4.1}{\spa1.3\spa3.4}\right)I_3(s_{13},m_t^2) \nonumber\\
&&+2m_t^2 \frac{\spb4.1}{\spb2.1}\left(\frac{\spa1.2\spb4.1}{\spa2.3^2}+\frac{\spa2.4\spb3.1}{\spa1.4\spa3.4}\right)I_3(s_{14},m_t^2) \nonumber\\
&&-2m_t^2\left(\frac{\spa1.2^3\spb2.1}{\spa1.3^2\spa1.4^2}-\frac{\spa1.2\spb3.1\spb4.1}{\spa1.4\spa1.3\spb2.1}\right)I_3(s_{12},m_t^2) 
\nonumber\\
&&-\frac{\spb3.1\spb4.1\spb4.3^2}{\spa1.2\spb2.1\spb3.2\spb4.2} \, .
\end{eqnarray}
The helicity amplitude for two negative helicity particles is 
\begin{eqnarray}
\mathcal{A}^{(1),m_t}_4(1_g^{-},2_g^{-},3_\gamma^{+},4_\gamma^{+} )&=&
m_t^2\left(\frac{\spa 1.2  \spb 4.3^2}{\spb 2.1}-2 m_t^2\frac{\spb 4.3^2}{\spb 2.1^2}  
\right)\bigg\{ I_4(s_{13},s_{12},m_t^2)+I_4(s_{14},s_{12},m_t^2)\bigg\}\nonumber\\&&
+\bigg\{\left(\frac{s_{14}s_{13} \spb4.3(
\spa1.4\spb3.2\spb4.1^2-\spa2.4\spb3.1\spb4.2^2)}{
2\spa1.2\spb2.1^4}
\right)\nonumber \\&& 
+m_t^2 \left(\frac{\spa 1.2  \spb 4.3^2}{\spb 2.1}-4 \frac{s_{14}s_{13}\spb4.3^2}{\spa1.2\spb2.1^3}
-2 m_t^2\frac{\spb 4.3^2}{\spb 2.1^2} \right)\bigg\}I_4(s_{13},s_{14},m_t^2) \nonumber\\&&
+\left(-\frac{(\spa1.4\spb3.2\spb4.1^2-\spa2.4\spb3.1\spb4.2^2)}{\spa3.4\spb2.1^2}
+4 m_t^2\frac{\spa1.2\spb3.2\spb4.1}{\spa3.4\spb2.1^2}\right)
\nonumber\\&& \times\bigg \{ s_{14} I_3(s_{14},m_t^2)+s_{13} I_3(s_{13},m_t^2)\bigg\}\nonumber\\&& 
+\left(\frac{\spa2.4^2(s_{13}-s_{14})\spb4.3}{\spa3.4^3\spb3.1^2}\right)\bigg\{I_2(s_{13},m_t^2)-I_2(s_{14},m_t^2)\bigg\}
\nonumber\\&& 
-\frac{\spa1.2^3\spb3.2\spb4.1}{\spa1.4\spa2.3\spb2.1^2} \, .
\end{eqnarray}     
Due to the Bose symmetry of these amplitudes, and trivial color ordering, all remaining helicity amplitudes can be
obtained by applying the appropriate re-orderings and conjugation operations to those listed above.
In the expressions above, the quantities $I_4(s,t,m_t^2)$, $I_3(s,m_t^2)$ and $I_2(s,m_t^2)$ represent the zero mass box,
one mass triangle and the bubble integral, respectively.  In all cases the internal propagators have a common mass, $m_t$.
In the notation of the QCDLoop library~\cite{Ellis:2007qk}, which we use to evaluate the integrals,
\begin{eqnarray}
&&I_4(s,t,m_t^2) \equiv I_4(0,0,0,s,t; m_t^2,m_t^2,m_t^2), \quad
I_3(s,m_t^2) \equiv I_3(s,0,0; m_t^2,m_t^2,m_t^2) \quad \nonumber\\
&&\mbox{and} \quad
I_2(s,m_t^2) \equiv I_2(s; m_t^2,m_t^2) \,.
\end{eqnarray}

\section{Rational amplitudes for $q\overline{q} g \gamma\gamma$: $n_F$ loops}
\label{app:nfrat}

One of the components of the N$^3$LO $\gamma\gamma$ contribution that we have computed consists of the one-loop
squared $q\overline{q}g\gamma\gamma$ amplitudes.  All of these amplitudes can be found in ref.~\cite{Campbell:2014yka},
with the exception of the $q^-\overline{q}^+g^+\gamma^+\gamma^+$ helicity assignment that does not contribute in that
calculation since it interferes with a vanishing tree-level amplitude.

We define the amplitude as in ref.~\cite{Campbell:2014yka}, namely,
\begin{eqnarray}
{A}^{(1)}(1^+_{q},2^-_{\overline{q}},3_g^+,4_{\gamma}^+,5_{\gamma}^+) = 
\sqrt{2} Q_i^2 \frac{\alpha_s}{2\pi} e^2 g_s (T^{a_3}_{i_1,i_2}) \mathcal{A}^{n_f}(1^+_{q},2^-_{\overline{q}},3_g^+,4_{\gamma}^+,5_{\gamma}^+) \, .
\end{eqnarray}
That is, we define our partial amplitude for a single loop of quarks of charge $Q_i$.
We note that all closed-loop diagrams in which the photon is radiated from the final state quark line
vanish either due to Furry's theorem (a single photon radiated from an external $q\overline{q}$) or proportionality
to tadpole diagrams (two photons emitted from external $q\overline{q}$). Our amplitude of interest is then given by 
\begin{eqnarray}
\mathcal{A}^{n_f}(1^+_{q},2^-_{\overline{q}},3_g^+,4_{\gamma}^+,5_{\gamma}^+)
=2 \, \frac{\spa2.3\spa4.5\spb4.1\spb5.3-\spa2.4\spa3.4\spb3.1\spb5.4}{\spa1.2\spa3.4\spa3.5\spa4.5\spb2.1} \, .
\end{eqnarray}
As must be the case for any amplitude that vanishes at tree-level, the one-loop amplitude is a rational function of the
external momenta.

\bibliographystyle{JHEP}
\bibliography{Gamgam}

\end{document}